\documentclass[a4paper,11pt]{article}
\pdfoutput=1
\usepackage[bottom]{footmisc}
\usepackage{jheppub}
\usepackage{calligra}
\usepackage{mathrsfs}
\usepackage{slashed}
\usepackage{graphicx}   %already in jheppub package
\usepackage{color}

\usepackage{feynmp}  
\usepackage{subcaption}
\newcommand{\CS}{\text{CS}}
\newcommand{\AdS}{\text{AdS}}

\newcommand{\CFT}{\text{CFT}}
\newcommand{\Mink}{\text{Mink}}

\newcommand{\Poincare}{\text{Poincare}}

\newcommand{\weq}{\underset{\text{Weyl}}{\sim}}
\newcommand{\slz}{\text{SL}(2,\textbf{Z})}

\title{AdS Asymptotic Symmetries  from CFT Mirrors}

\author[a]{Rashmish K.~Mishra,}
\author[b]{Arif Mohd}
\author[b]{and Raman Sundrum}

\affiliation[a]{INFN, Pisa, Italy and Scuola Normale Superiore, Piazza dei Cavalieri 7, 56126 Pisa, Italy}
\affiliation[b]{Maryland Center for Fundamental Physics, Department of Physics, University of Maryland, College Park, MD 20742, USA}

\emailAdd{rashmish@pi.infn.it}
\emailAdd{arif7de@gmail.com}
\emailAdd{raman@umd.edu}

\abstract
{
We study Kac-Moody asymptotic symmetries and memory effects in $\AdS_4^{\Poincare}$ gauge theory and (when accompanied by 4D gravity) in its holographic CFT$_3$ dual. While such infinite-dimensional symmetries are absent in standard asymptotic analyses of $\AdS_4$, we show how they arise with  alternate AdS boundary conditions. In the 3D holographic description, these alternate boundary conditions correspond to a modified $\widetilde{\CFT}_3$ obtained by Chern-Simons gauging of the CFT$_3$ dual defined by standard boundary conditions, so that Kac-Moody symmetries then follow from the familiar Chern-Simons/Wess-Zumino-Witten correspondence. Apart from their own intrinsic interest, in abelian $\AdS_4$ gauge theories these alternate boundary conditions are equivalent to standard boundary conditions imposed on electric-magnetic dual variables. In the holographic description this corresponds to 3D ``mirror" symmetries connecting the original and modified CFTs. Further, in both abelian and non-abelian theories we show that the alternative/$\widetilde{\CFT}_3$ theory emerges at leading order in large Chern-Simons level from the correlators of the standard theory, upon incorporating large-wavelength limits in the holographically emergent dimension. We point out similarities and differences between 4D AdS and Minkowski gauge theories in their asymptotic symmetries, ``soft'' limits and memory effects.
}

\begin{document}
\begin{flushright}
UMD-PP-018-06 \\
\end{flushright}
\maketitle
\flushbottom

\section{Introduction}
\label{sec:intro}

In gravitational and gauge theories, asymptotic symmetries (AS) are a global remnant of large diffeomorphisms and gauge transformations which act non-trivially on physical data at spacetime infinity. The classic example of infinite-dimensional AS, and in many ways the best understood and applied, is that of (quantum) General Relativity (GR) in asymptotically 3D Anti-de Sitter (AdS$_3$) spacetime.  The analysis of Brown and Henneaux~\cite{Brown:1986nw} uncovered Virasoro symmetries which presaged, and were ultimately elegantly incorporated into, the AdS$_3$/CFT$_2$ correspondence, translating into the implications of 2D conformal invariance and unitarity. The Virasoro structure and central charges, with modular invariance,  led to a precise microscopic account~\cite{Strominger:1997eq} of the Bekenstein-Hawking entropy of AdS$_3$ Schwarzchild black holes, dual to the CFT$_2$ Cardy formula~\cite{Cardy:1986ie}. There is an ongoing program of exploiting this symmetry structure  to address more detailed aspects of  black hole information puzzles~\cite{Fitzpatrick:2016mjq, Fitzpatrick:2016ive}. In a similar vein to these gravitational asymptotic symmetries, 3D Chern-Simons (CS) gauge theories display infinite-dimensional Kac-Moody (KM) asymptotic symmetries with central extensions, reflecting 2D Wess-Zumino-Witten  (WZW) current algebras via the technically simpler CS/WZW correspondence~\cite{Witten:1988hf, Elitzur:1989nr, Witten:1991mm, Gukov:2004id}.

In higher dimensions the situation is intriguing, but less well understood. The primordial example is provided by the infinite-dimensional BMS ``supertranslations'' of GR in asymptotically 4D Minkowski spacetime (Mink$_4$)~\cite{Bondi:1962px, Sachs:1962wk}, later extended to include Virasoro-type ``superrotations''~\cite{Barnich:2009se, Barnich:2011ct}, and Kac-Moody asymptotic symmetries from 4D gauge theory~\cite{Strominger:2013lka, He:2014cra, He:2015zea, Kapec:2015ena}. However, the symmetry algebras have appeared without central extensions, ordinarily required by unitarity in lower-dimensional contexts. There are new deep aspects in 4D, unifying asymptotic symmetries with soft limits of gravitons and gauge bosons, and with in-principle physical gravitational and gauge ``memory'' effects (see Ref.~\cite{Strominger:2017zoo} for a review and extensive list of references).  There are also hopes of applying AS to help understand black hole information~\cite{Hawking:2016sgy, Hawking:2016msc, Strominger:2017aeh, Carney:2017jut, Carney:2017oxp}, although this is still under debate~\cite{Bousso:2017dny,Mirbabayi:2016axw, Gabai:2016kuf, Gomez:2017ioy, Donnelly:2017jcd, Bousso:2017rsx}. The asymptotic symmetries can be shown to derive from 2D current algebras ``living'' on the celestial sphere,  but it is unclear what the precise connection is between this structure and some form of holography in Minkowski spacetime. One hint comes from an intermediate step between 4D and 2D:  the soft limit of gravitational and gauge fields renders them effectively 3-dimensional, in a more nuanced generalization of the trivial loss of the time dimension in the {\it static} limit. In particular, some of the soft fields take the form of 3D GR and CS~\cite{Cheung:2016iub}, with close ties to the AdS$_3$/CFT$_2$ and CS/WZW correspondences~\cite{Witten:1988hf, Elitzur:1989nr, Witten:1991mm, Gukov:2004id}.

In order to explore the connection of  4D asymptotic symmetries to holography, Ref.~\cite{Mishra:2017zan} turned to the study of asymptotic symmetries in (portions of) AdS$_4$, taking advantage of the well-established $\AdS_4/\CFT_3$ correspondence. In this context, there is a natural way to include 3D (conformal) GR and CS, by simply having them gauge the holographic CFT$_3$ at the outset.  Applying 3D (conformal) GR and CS ($+$ CFT$_3$ ``matter") analyses
 then yields a set of infinite-dimensional asymptotic symmetries {\it with central extensions}.  Even in the limit in which the external 3D GR and CS fields decouple from CFT$_3$, these asymptotic symmetries symmetries remain, but {\it losing their central extensions} as the price for restricting to CFT correlators with a well-defined decoupling limit. The resulting asymptotic symmetries closely parallel the supertranslation, superrotation and Kac-Moody asymptotic symmetries of Mink$_4$.

In this paper, we continue the study of asymptotic symmetries in the context of $\AdS_4/\CFT_3$.
 We restrict our attention to gauge theory in the Poincare patch of AdS$_4$ for technical and conceptual simplicity, with 4D GR only an incidental presence needed for duality with CFT$_3$. Within this framework, we will identify different but interconnected ways in which Kac-Moody asymptotic symmetries arise. Most directly we extend the approach of Ref.~\cite{Mishra:2017zan} to the Poincare patch, with CS-gaugings of the holographic CFT defining new $\widetilde{\CFT}$s, and the canonical CS structure leading to Kac-Moody asymptotic symmetries with finite central extensions. The AdS dual of the modified $\widetilde{\CFT}$  shares the same 4D dynamics as the AdS dual of the original CFT, but with the former
 having an alternate set of AdS boundary conditions~\cite{Witten:2003ya}(particular to 4D). This is key to evading no-go arguments~\cite{Ashtekar:1999jx, Papadimitriou:2005ii} for infinite-dimensional asymptotic symmetries in $\AdS_{d> 3}$.

In the case of {\it abelian} gauge/global symmetries of AdS$_4$/CFT$_3$, we can make a stronger statement because the original CFT and the $\widetilde{\CFT}$s are connected by $\slz$ ``mirror" symmetry~\cite{Witten:2003ya}.  From the AdS$_4$ viewpoint, this $\slz$ is associated to electric-magnetic duality, which relates the standard boundary conditions to alternate boundary conditions. In this sense, Kac-Moody asymptotic symmetries structure 
already resides in the standard AdS$_4$/CFT$_3$ construction, albeit applied in suitable electric-magnetic/mirror dual variables.
 
For both abelian and non-abelian theories, there is another way in which we will show that the standard AdS$_4$/CFT$_3$  theory contains the ``seeds" of the alternate/$\widetilde{\CFT}$ theory, namely by 
 taking gauge-boson long-wavelength limits in the holographically emergent dimension within $\partial \AdS_4$ correlators.  We show that this ``holographic soft limit" of the standard theory yields the correlators and Kac-Moody asymptotic symmetries of the alternate theory to leading order in the CS level,  closely matching and adding physical significance to  the decoupling limit AS analysis of  Ref.~\cite{Mishra:2017zan}.  
Paralleling the connections in $\Mink_4$ between asymptotic symmetries, soft limits and memory effects, we will show  in $\AdS_4$ abelian gauge theory that  the KM asymptotic symmetries and holographic soft limits are closely connected to ``magnetic" gauge memory effects.

The paper is organized as follows. In Section~\ref{sec:ads4-poincare}, we introduce gauge theory in the Poincare patch of $\AdS_4$, standard and alternate boundary conditions, and their holographic translations in terms of $\CFT_3$ and $\widetilde{\CFT}_3 \equiv \CS + \CFT_3$, respectively.  In Section~\ref{sec:KM-from-MS} we derive the Kac-Moody asymptotic symmetries of the alternate $\AdS_4/\widetilde{\CFT}_3$ theory from its canonical CS structure. In Section~\ref{sec:EMduality-from-MirrorDuality}, we restrict to abelian theories and 
point out the passage from standard $\AdS_4/\CFT_3$ to alternate $\AdS_4/\widetilde{\CFT}_3$, and hence Kac-Moody asymptotic symmetries,  via electric-magnetic/mirror duality. In Section~\ref{sec:MirrorDual-from-SoftLimit-abelian}, we derive another passage from standard $\AdS_4/\CFT_3$ to alternate $\AdS_4/\widetilde{\CFT}_3$ in abelian theories, this time by introducing the ``holographic soft limit" in its simplest form. In Section~\ref{sec:MirrorDual-from-SoftLimit-nonAbelian} we generalize this soft limit analysis to non-abelian gauge theories in $\AdS_4$, involving more careful treatment of multiple soft external lines.
 In Section~\ref{sec:memory} we describe (abelian) magnetic memory effects in standard $\AdS_4/\CFT_3$ and give their holographic interpretation and connections to Kac-Moody asymptotic symmetries structure and soft limits. We provide our conclusions in section~\ref{sec:discussion}, including several parallels and contrasts between the $\AdS_4$ and $\Mink_4$ asymptotic symmetry analyses.

\section{$\AdS_4$ Gauge Theory, Boundary Conditions and Holography}  
\label{sec:ads4-poincare}

We describe the Poincare patch of $\AdS_4$ by coordinates $X^M \equiv (t,x,y,z)$ and metric, 
\begin{align}
ds_{\AdS_4}^2 & = \frac{dt^2-dx^2-dy^2-dz^2}{z^2}, ~ z>0,
\end{align}
where we work in units of the $\AdS$ radius of curvature.
Its boundary, $\partial \AdS_4 \equiv \Mink_3$, is at $z=0$, with 3D coordinates $x^{\mu} \equiv (t,x,y)$. 
We consider AdS dynamics of the form,
\begin{align}
\mathcal{L}_{\AdS_4}
&=
-\frac{1}{2g^2}\,\text{Tr}\,\mathcal{F}_{MN} \mathcal{F}^{MN} 
+ \frac{\theta}{16\pi^2} \,\text{Tr}\, \mathcal{F}_{MN} \widetilde{\mathcal{F}}^{MN}
+\mathcal{A}_M^a \mathcal{J}^{M\:a} + \cdots\:,
\label{eq:AdS-dynamics}
\end{align}
where $\mathcal{A}_M \equiv \mathcal{A}_M^a t^a$ is a 4D gauge field with field strength $\mathcal{F}_{MN} \equiv \mathcal{F}_{MN}^a t^a$, $\mathcal{J}_M^a$ is the 4D current due to gauge-charged matter, $t^a$ are the generators of gauge group, normalized as $\text{Tr}\, t^a t^b = \delta^{ab}/2$, and the ellipsis includes the 4D matter Lagrangian as well as 
 4D quantum gravity.  We will not explicitly need the details of quantum gravity in this paper, but with it the $\AdS_4$ theory has a $\CFT_3$ holographic dual on Mink$_3$, which we will invoke (see Refs.~\cite{Aharony:1999ti, Sundrum:2011ic} for a review). 

\subsection{Standard ``Dirichlet'' Boundary Conditions} 
\label{sec:ads4-poincare-dirichlet}
The standard $\AdS_4$ boundary condition (b.c.) is 
\begin{align}
\mathcal{A}_\mu^a(x^\nu,z) \xrightarrow[z \rightarrow 0]{} A_\mu^a(x^\nu)\:,
\end{align}
where $A^a_{\mu}(x^\nu)$ is the source for the dual $\CFT_3$ conserved global current, $J_{\mu}^a(x^\nu)$. 
The 4D $\theta$-term introduces a subtlety, seen by the decomposition,
\begin{align}
\theta = \bar{\theta} + 2\pi \kappa,\:\:\bar{\theta} \in [0,2\pi),\:\: \kappa \in \mathbb{Z}.
\label{eq:theta-decomposition}
\end{align}
4D bulk physics only depends on the angle $\bar{\theta}$ as usual. For simplicity, in this paper we restrict attention to $\bar{\theta} =0$. However,  given the total derivative nature of the $\theta$-term, 
$\kappa$ survives as a $\partial \AdS_4$ action for the source $A_\mu$, 
\begin{align}
\mathcal{L}_{\Mink_3}  = \mathcal{L}_{\CFT_3} + A_\mu^a J^{\mu\, a} 
+ \frac{\kappa}{4\pi} \: \epsilon^{\mu\nu\rho} \:
\text{Tr} \left(
A_\mu \partial_\nu  A_\rho 
+ \frac23 A_\mu A_\nu A_\rho
\right).
\label{eq:CS-action}
\end{align}
This gives extra contact terms, consistent with 3D conformal invariance, in multi-current correlators at coincident points~\cite{Witten:2003ya}. For example, 
\begin{align}
\left< T\left\{J_\mu^a(x)J_\nu^b(x')\cdots\right\} \right> 
\:\:\supset \:\: 
\kappa \: \epsilon_{\mu\nu\rho}\delta^{ab}
\partial^\rho\delta^3(x-x')
\left< \cdots \right>.
\end{align}
For vanishing source, $A = 0$, the boundary condition takes the ``Dirichlet'' (D) form $\mathcal{A}_\mu(x^\nu,z)  \xrightarrow [z\rightarrow 0]{} 0$, or more gauge-invariantly,
\begin{align}
\mathcal{F}_{\mu\nu}^a (x^\nu, z) \xrightarrow [z \rightarrow 0]{} 0\:,
\label{eq:Dbc}
\end{align}
since the 3D dual description is also gauge-invariant if we transform the source $A_\mu$ as a background 3D gauge field.

\subsection{CS-gauged CFT$_3$ and Alternate Boundary Conditions}
\label{sec:ads4-poincare-neumann}

We define a modified $\widetilde{\CFT}_3$ by simply elevating the source $A_\mu$ above to a fully dynamical field with the same action, Eq.~\eqref{eq:CS-action}. The $\kappa$ terms no longer represent contact terms for global current correlators of $\CFT_3$, but rather a CS action for $A_\mu$, which then gauges the $\CFT_3$ current $J_\mu$. Schematically, 
$\widetilde{\CFT}_3 = \CS + \CFT_3$. 

The $\AdS_4$ dual of $\widetilde{\CFT}_3$ is given by the same bulk dynamics as for the original $\CFT_3$ but with an alternate boundary condition~\cite{Witten:2003ya}. A large set of gauge-invariant boundary conditions respecting the $\AdS_4$ isometries (3D conformal invariance) exist because one can replace the ``Dirichlet" vanishing of ${\cal F}_{\mu \nu}$ at the boundary by vanishing of a more general linear combination of ${\cal F}_{\mu \nu}$ and $\widetilde{\cal F}_{\mu \nu}$. We see that the CS equations of motion corresponding to the action of Eq.~\eqref{eq:CS-action} is matched by alternate boundary condition of the form, 
\begin{equation} 
\frac{\kappa}{2\pi} {\cal F}_{\mu \nu} + \frac{1}{g^2}\widetilde{\cal F}_{\mu \nu} \xrightarrow [z \rightarrow 0]{} 0,
\label{eq:DandNbc}
\end{equation}
because of the standard holographic matching
\begin{equation}
2
\widetilde{\cal F}_{\mu \nu} \equiv  \epsilon_{\mu \nu \rho z} {\cal F}^{z \rho}  \xrightarrow [z \rightarrow 0]{} 
2g^2\,\epsilon_{\mu \nu \rho} J^{\rho}.
\end{equation}

In the simplest case, $\kappa = 0$, the alternate boundary condition is just a gauge invariant version of 
 ``Neumann'' (N) boundary condition: 
\begin{align}
2
\widetilde{\mathcal{F}}_{\mu\nu} 
\equiv
\epsilon_{\mu\nu\rho z}\mathcal{F}^{z \rho}  
\xrightarrow [z \rightarrow 0]{} 0\:,
\label{eq:Nbc}
\end{align}
as is clear in axial gauge $\mathcal{A}^a_z = 0$, 
\begin{align}
\mathcal{F}_{z \rho} = \partial_z \mathcal{A}_\rho  \xrightarrow [z \rightarrow 0]{} 0\:.
\label{eq:Nbc-axialGauge}
\end{align}

\section{Kac-Moody AS from CS Structure}
\label{sec:KM-from-MS}
\begin{figure}[h]
\centering
\includegraphics[width=.7\linewidth]{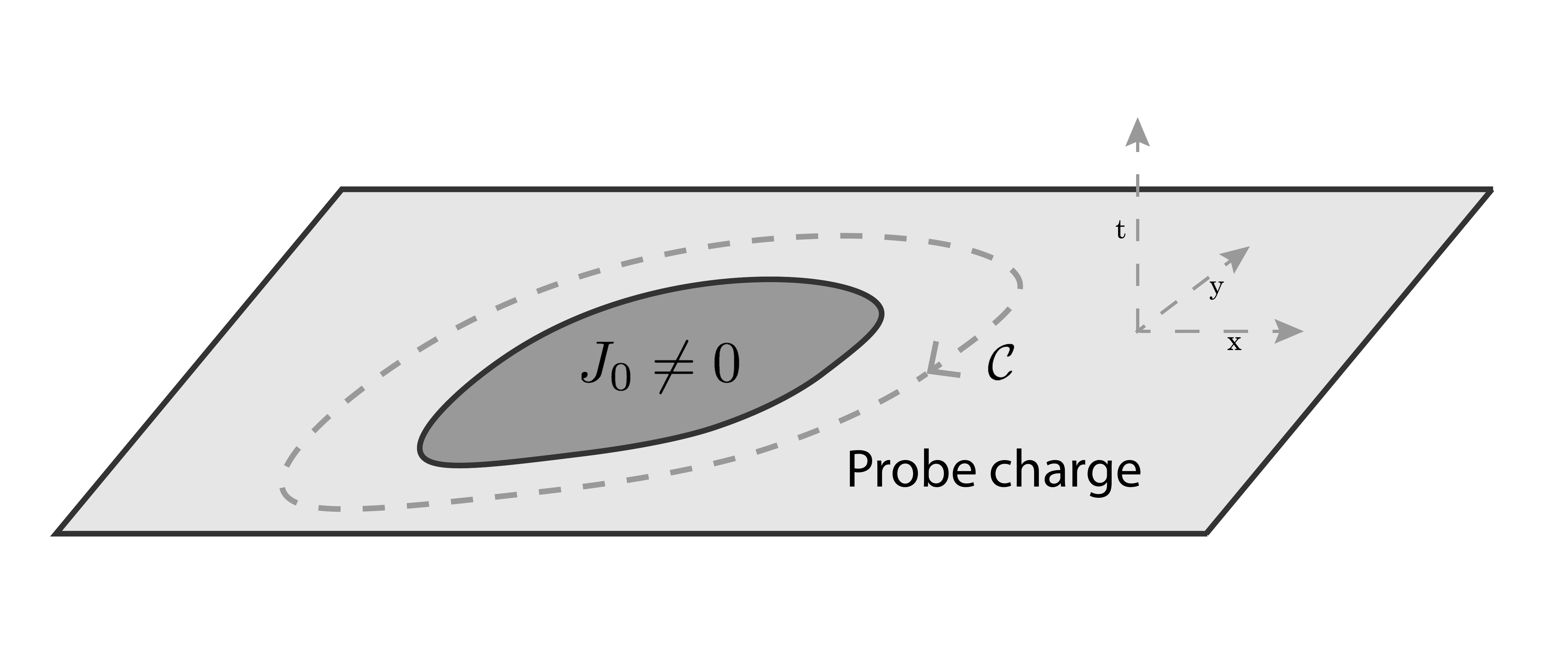}
\caption{\small{Non-trivial Wilson loops ${\cal C}$ enclosing charge density, giving rise to Aharanov-Bohm type effects on test charges.}}
\label{fig:CS-as-AB-effect}
\end{figure}

In this section we consider the above AdS$_4$ gauge theory ($+$ quantum gravity) with alternate boundary condition, or equivalently in 3D, $\widetilde{\CFT}_3 \equiv$ CS $+$ CFT$_3$, with level $\kappa$. 
3D CS gauge theory coupled to matter (provided here by $\CFT_3$)  describes relativistic (non)-abelian Aharanov-Bohm type effects between separated charges (e.g. see Ref.~\cite{Tong:2016kpv} for a review), thereby providing charged matter with quantum ``topological hair''. This is manifest already in the CS Gauss Law constraint ($A^a_0$ equation of motion),
\begin{align}
\frac{\kappa}{2\pi}\, F_{xy}^a = J_0^a\:,
\end{align}
where $F_{\mu\nu}^a$ is the field strength of $A$. Outside the support of the charge density $J_0$, $F_{xy} = 0$, but spatial Wilson loops (as seen by test charges) here
are non-trivial when enclosing charge $J_0$, as in Fig.~\ref{fig:CS-as-AB-effect}. 

Related to the topological nature of their Aharanov-Bohm effects, CS structure on 3D spacetimes with a 2D boundary can be mapped to WZW 2D current algebras, exhibiting Kac-Moody asymptotic symmetries at the 2D boundary~\cite{Witten:1988hf, Elitzur:1989nr, Witten:1991mm, Gukov:2004id}. In the present context however,  CS lives on $\Mink_3$, with no finite 2D boundary. But from the canonical viewpoint the state wavefunctional, $\Psi$,  at some fixed time, say $t=0$, {\it does} exhibit Euclidean signature WZW/KM structure on the spatial $x-y$ plane at that time, the relevant Ward identities supplied by Gauss' Law~\cite{Witten:1988hf}. One can think of  $\Psi(t=0)$ as given by a $\CS$ + $\CFT_3$ path integral on the earlier half of $\Mink_3$,  $t < 0$,  a spacetime with 2D boundary $t=0$. 

\subsection{Gauss Law constraints on canonical CS fields}
\label{subsec:GaussLawConstraint}

To review this, we introduce complex coordinates, 
\begin{equation}
u \equiv x + i y, ~  ~ \bar{u} \equiv x - i y,
\end{equation}
in which Gauss' Law ($A_0$ equation of motion) reads
\begin{align}
\left(
\partial_{\bar{u}} j^a - 2i\kappa\,\partial_u A_{\bar{u}}^a - f^{abc} j^b A_{\bar{u}}^c
\right)
\Psi[A_{\bar{u}}] 
=
2\pi J^a_0  \Psi[A_{\bar{u}}].
\label{eq:gauss-law}
\end{align}
To explain our notation, from Eq.~\eqref{eq:CS-action} we see from the CS Lagrangian that (after integrating out $A_0^a$)  $A_u$ and $A_{\bar{u}}$ are canonically conjugate. Here, we choose to  work in $A_{\bar{u}}$ field-space, and denote a (non-canonically normalized, for later convenience) conjugate field-momentum by
\begin{align}
j^a(u,\bar{u})
\equiv 
i\pi \frac{\partial \mathcal{L}_{\CS}}{\partial \dot{A}^a_{\bar{u}}}
= 2i\kappa \, A_u^a.
\end{align}
The wavefunctional $\Psi$ is taken to depend on $A_{\bar{u}}$  (coherent state representation) and the CFT fields. At the quantum level the conjugate field-momentum is then given by 
\begin{equation}
j^a(u,\bar{u}) = i\pi \frac{\delta}{\delta A_{\bar{u}}^a(u, \bar{u})},
\end{equation}
The quantum Gauss' Law has the form of  a functional differential equation that effectively determines the $A_{\bar{u}}$-dependence of the wavefunctional in terms of the matter CFT state.

\subsection{Holomorphic  2D WZW current and KM symmetry from CS}
\label{subsec:2DWZWCurrent}

For simplicity, we begin by exploring $\Psi$ at $A_{\bar{u}} = 0$ and for the special case of the CFT state consisting only of pointlike disturbances at $t=0$, 
\begin{align}
\Psi \propto \prod_n \mathcal{O}_n(u,\bar{u})\left|0\right>\:,
\end{align}
where the $\mathcal{O}$ are local operators. We discuss more general $A_{\bar{u}}$ below, and more general CFT
states in the next subsection.

For the special state above, Gauss' Law reduces to 
\begin{align}
\partial_{\bar{u}} j^a(u,\bar{u})
\Psi [ A_{\bar{u}} = 0 ]\:
=
2\pi\,\sum_{\alpha=1}^n 
T_{(\alpha)}^a 
\delta^2 (u - u_\alpha) 
\Psi [ A_{\bar{u}} = 0 ]\: ,
\label{eq:ward-identity}
\end{align}
where $T^a_{(\alpha)}$ is the representation of the (non-)abelian generator acting on the particular local CFT operator ${\cal O}_{\alpha}(u_{\alpha}, \bar{u}_{\alpha})$, giving its charge. This equation can be integrated\footnote{We are assuming the wavefunctional is a well-behaved function of $A_{\bar{u}}$ at infinity, so that we do not have to include an analytic function of $u$ as integration constant in RHS of Eq.~\eqref{eq:IntegratedGaussLaw}.} to give
\begin{align}
j^a(u,\bar{u})\Psi[A_{\bar{u}} = 0 ]
=
\sum_\alpha
\frac{T_{(\alpha)}^a}{u-u_\alpha}\Psi[A_{\bar{u}} = 0]\:,
\label{eq:IntegratedGaussLaw}
\end{align}
using the identity $\partial_{\bar{u}} \left(1/(u - u_{\alpha})\right) = 2\pi\,\delta^2(u - u_{\alpha})$. From this we can then extract a 2D ``OPE'', matching that of a standard holomorphic WZW current with a charged operator in 2D Euclidean field theory (e.g. see Ref.~\cite{DiFrancesco:1997nk} for a review),
\begin{align}
j^a(u,\bar{u})\mathcal{O}_\alpha (u_\alpha, \bar{u}_\alpha) 
\:\: \xrightarrow[u\rightarrow u_\alpha]{} \:\: 
\frac{T_{(\alpha)}^a\mathcal{O}_\alpha(u_\alpha, \bar{u}_\alpha)}{u-u_\alpha}.
\end{align}

Next, we begin with non-vanishing $A_{\bar{u}} $ and act on Gauss' Law with the operator $j^b(u', \bar{u}') \equiv i\pi \delta/\delta A^b_{\bar{u}}(u', \bar{u}')$, and only then set $A_{\bar{u}} = 0$:
\begin{align}
\left[
\kappa \: \partial_u \delta^2(u-u')\delta^{ab}
+ \frac{1}{2\pi} \partial_{\bar{u}} j^a j^{b'} 
-\frac{i}{2} f^{abc} \delta^2(u-u') j^c
\right]
\Psi
= j^b(u',\bar{u}')J_0^a(u,\bar{u})\Psi\:.
\end{align}
We consider $u$ away from any CFT local operators at $u_{\alpha}$ (within $\Psi$), so the right-hand side is non-singular in $u - u'$. The left-hand side can again be integrated, using the identity $-\partial_{\bar{u}}\left( 1/(u - u')^2\right) = \partial_{\bar{u}} \partial_u \left(1/(u - u')\right) = 2\pi\,\partial_u \delta^2(u - u')$, to give the $j j'$ OPE, 
\begin{align}
j^a(u,\bar{u}) j^b(u',\bar{u}')
\:\: \xrightarrow[u\rightarrow u']{} \:\:
\frac{\kappa}{\left(u - u'\right)^2}\delta^{ab} + \frac{if^{abc}}{2(u-u')}j^c\:.
\label{eq:jj-OPE-Abelian}
\end{align}
Choosing $u'=0$ the 2D holomorphic current can be expanded in a Laurent expansion of KM charges 
\begin{align}
j^a(u) \equiv \sum_m \frac{Q_m^a}{u^{m+1}}\:,
\end{align}
Plugging this into the OPE and interpreting the result in standard 2D Euclidean radial quantization gives the KM symmetry algebra,  
\begin{align}
\left[
Q_m^a, Q_n^b
\right]
= \kappa \: m \:\delta^{ab}\: \:\delta_{m,-n}
+ if^{abc}\: Q^c_{m+n}\:,
\end{align}
where the central extension is provided by the CS level $\kappa$. 

Via $\AdS_4/\CFT_3$ duality, we then conclude that with alternate boundary condition, Eq.~\eqref{eq:DandNbc}, 
this CFT derivation of the Kac-Moody algebra structure translates to 
$\AdS_4$  gauge theory. So far our derivation focused on the special CFT state with all charged local operators acting on the vacuum at the same time, $t=0$, dual to all charged lines in $\AdS_4$ arriving at the boundary at the same time $t=0$. Below, we consider more general CFT/AdS states.

\subsection{General CFT states and non-holomorphicity of WZW current}
\label{subsec:generalCFTstates}

More typical CFT states cannot be described by purely local disturbances of the vacuum, created by just local operators at $t=0$. Instead, we can think of them as follows. If we consider the CFT to have a large-$N$ type gauge theoretic structure, it will contain CFT-gauge charged ``quarks" also transforming under a global symmetry of the CFT, which is then gauged by CS. The state at $t=0$ will consist of CFT-gauge singlet combinations of these 3D ``quarks" and ``gluons", but the quarks in a minimal CFT-singlet will typically not all be localized at a  single point, but rather dispersed to some extent in 2D space. 
From this fundamental $\CFT_3$ perspective, our construction of $j$ will still be a holomorphic current,  with simple poles at the locations of the 3D quarks at $t=0$, and the entire KM algebra and symmetry structure via Gauss' Law still follows straightforwardly.

However, from the $\AdS_4$ dual perspective individual CFT quarks are not explicitly described, rather the 4D description is an effective ``hadronic" description of the different CFT-gauge-singlet combinations of 3D quarks and gluons, 
 in terms of which we only see a ``smeared" continuum approximation to the fundamentally pointlike quark CS-charges, with $J_0$ taking the form of the boundary limit of the 4D transverse electric field. 
 Local CFT/boundary operators can still be used to interpolate the more general states, but they must be allowed to act {\it before} $t=0$ so that their disturbance of the vacuum can spread by $t=0$. This is dual to  4D particles created at the boundary at early times having moved off into the bulk  of $\AdS_4$ by $t=0$.
 
We illustrate the nature of this smearing in the case of abelian CS symmetry. The discrete sum over CS-charge locations  in Eq.~\eqref{eq:ward-identity} is more generally replaced by the charge density $J_0$ as in Eq.~\eqref{eq:gauss-law}, so that $j$  in Eq.~\eqref{eq:IntegratedGaussLaw} is replaced by a ``smeared" integral over poles, 
 \begin{equation}
 j = \int d^2 u' \frac{J_0(u', \bar{u}')}{u - u'},
 \end{equation}
rather than the discrete sum of poles that is more familiar from standard CS/WZW contexts. Nevertheless, we know from the CFT quark perspective that the KM symmetry structure is fully intact for general states. 
Even at the smeared level of description, the meaning of the KM charges can be discerned. For example, if we consider a state at $t=0$ with some finite region of support  for $J_0$, then $j$ is holomorphic outside this region. 
If the support of $J_0$ excludes the origin, we can expand for small $u$, 
\begin{equation}
j = - \sum_{n\geq0} \int d^2 u' \frac{J_0(u', \bar{u}') u^n}{u'^{n+1}},
\end{equation} 
 corresponding to KM charges as moments of the charge distribution,
 \begin{equation}
Q_n = - \int d^2 u' \frac{J_0(u', \bar{u}')}{u'^n}, ~ n < 0.
\end{equation} 
We can also expand for large $u$ compared to the support of $J_0$, 
\begin{equation}
j = \sum_{n\geq0} \int d^2 u' \frac{J_0(u', \bar{u}') u'^n}{u^{n+1}},
\end{equation} 
 thereby identifying effective KM charges,
 \begin{equation}
Q_n = \int d^2 u' J_0(u', \bar{u}') u'^n, ~ n \geq 0.
\end{equation}
    
In later sections we will discuss ``smeared" KM structure and associated memory effects in the context of  $\partial \AdS_4$ correlators  with standard Dirichlet boundary conditions,
which more closely parallel features of the  Mink$_4$ S-matrix and memory effects. Nevertheless,  the above features of KM structure from the canonical wavefunctional viewpoint for (the holographic dual of) alternate boundary conditions are already somewhat reminiscent of  $\Mink_4$. 
 The 2D KM current construction in $\Mink_4$ gauge theory, has simple poles at angular locations of charged particles arriving at lightlike infinity, ${\cal I}^+$. But here too this simple pole structure can be smeared out if 
the charged particles instead arrive at timelike infinity~\cite{Strominger:2017zoo, Kapec:2015ena, Campiglia:2015qka}.  
However, in $\Mink_4$  the final destination of charged particles is determined by their 4D mass, massless charges automatically arriving at ${\cal I}^+$ and massive charges at timelike infinity. In this sense, the simple pole structure in $\Mink_4$ is more readily arranged, by restricting to a final state with only massless charges. By contrast in $\AdS_4$, the restricted states at $t=0$ yielding simple pole structure do not follow automatically by restricting the 4D particle species/masses of the final state.

Amusingly, the holographic perspective reveals that there is indeed a correlation between the mass of charges and the robustness of the simple pole structure of the 2D KM currents, but the correlation is given in terms of 3D holographic masses! Furthermore, it is for the massive case that the simple pole structure is robust and for the massless case that it is not. In CS theories with  massive 3D charged species, the restriction to states with a few pointlike charged excitations at $t=0$ is automatic given a  finite energy ``budget", yielding simple-pole structure of $j$ generally. 
 But a $\CFT_3$ consists  instead of 3D-massless  (and strongly-coupled) ``quarks" as discussed above, so a typical state is a collection of indefinite numbers of these ``quarks".

\section{AS from  4D Electric-Magnetic Duality/3D Mirror Symmetry}
\label{sec:EMduality-from-MirrorDuality}

We have seen that alternate AdS$_4$ boundary condition, dual to the modified $\widetilde{\CFT}_3$, explicitly contains CS and hence CS/WZW-related KM structure. But this analysis seems to exclude the case of standard AdS$_4$ boundary condition, dual to the isolated 
original $\CFT_3$.   The remainder of this paper is devoted to showing different senses in which even this original unmodified theory does connect to Kac-Moody asymptotic symmetries. In this section, we will show that in the case of {\it abelian} AdS$_4$ gauge symmetry there is a full CS and Kac-Moody asymptotic symmetries structure arising from standard boundary condition, when these are imposed on
the 4D gauge theory in suitable electric-magnetic dual variables.  At the holographic level, this shows how the standard and modified CFTs transform into one another via 3D mirror symmetries.

The most familiar form of  
electric-magnetic duality arises from the invariance of pure Maxwell theory under
\begin{align}
\mathcal{F} \rightarrow \widetilde{\mathcal{F}},\:
\widetilde{\mathcal{F}} \rightarrow -\mathcal{F}\:.
\end{align}
More precisely, in the presence of charged matter it is described by a discrete duality transformation, $S$, which acts on states with electric charge $n g$ and magnetic charge $2 \pi m/g$  (where $n,m$ are integers for Dirac quantization) according to
\begin{equation}
S(n,m) = (m, -n).
\end{equation}
From the viewpoint of the 4D magnetic dual gauge field, $\widetilde{A}_M: ~ \widetilde{F}_{MN} = \partial_M \widetilde{A}_N - \partial_N \widetilde{A}_M$, the roles of the ``standard'' D and ``Neumann'' N boundary conditions are exchanged, as is clear from their gauge-invariant forms, Eq.~\eqref{eq:Dbc}, and Eqs.~\eqref{eq:Nbc}, \eqref{eq:Nbc-axialGauge}. 
That is, $
D \equiv \widetilde{N}, ~ N \equiv \widetilde{D}$.

Electric-magnetic duality extends to a full $\slz$, generated by $S$ and $T$, where $T$ corresponds to the shift  in the CP-violating parameter $\theta \rightarrow \theta + 2\pi$, another invariance of the bulk 4D physics.
 Witten has pointed out that general shifts in $\theta$ induce shifts in the spectrum of electric charges of states with non-zero magnetic charge. For the $(2\pi){\rm  integer}$ shift of $T$ this Witten effect~\cite{Witten:1979ey} corresponds to
\begin{equation}
T(n,m) = (n + m, m).
\end{equation}
In this way, $\slz$ duality exchanges ordinary electric charges with more general dyonic charges $(n,m)$. 

As we saw for the $S$ transformation above, the 
 AdS boundary conditions are not invariant under the  more general $\slz$ transformations, since they pick out the particular type of $(n,m)$ charge whose gauge field is given Dirichlet boundary condition, thereby defining 
 the global current of the dual CFT. The
 standard boundary condition picks out ordinary electric charges $(1,0)$ of course. 
 For a general $(n,m)$ the boundary conditions involve an obvious linear combination of the Dirichlet and Neumann boundary conditions, 
\begin{align} 
g n \mathcal{F}_{\mu \nu} 
+ 
\frac{2 \pi m}{g}\widetilde{\mathcal{F}}_{\mu \nu} 
\xrightarrow[z\rightarrow 0]{}  0.
\end{align}
$\slz$ thereby incarnates as 3D mirror symmetry, transforming between the different CFTs given by these different boundary conditions. 

For example, if we first apply the $TS$ transformation to the 4D gauge theory and {\it then} impose standard boundary conditions, we get Dirichlet boundary condition applied to the gauge field that couples to  $TS(1,0) = (-1,-1)$ charges, 
\begin{align} 
g \mathcal{F}_{\mu \nu} 
+ 
\frac{2 \pi }{g}\widetilde{\mathcal{F}}_{\mu \nu} 
\xrightarrow[z\rightarrow 0]{}  0.
\end{align}

From the discussion of subsection~\ref{sec:ads4-poincare-neumann}, we see that this corresponds to a CS gauging of the original $\CFT_3$, with level $\kappa = 1$. 

In this way, $\slz$ equates the standard boundary conditions of $\AdS_4$ gauge theory with alternative boundary conditions, which then manifest Kac-Moody asymptotic symmetries as described earlier. 
\section{Alternate/$\widetilde{\CFT}$ Correlators from ``Holographic Soft Limit"}
\label{sec:MirrorDual-from-SoftLimit-abelian}

We now turn to the sense in which the standard $\AdS_4$ Dirichlet boundary condition, dual to $\CFT_3$ in isolation, has implicit CS structure and AS in the original ``electric'' variables once we include a natural $\AdS^{\Poincare}$ generalization of the notion of ``soft limit", applying whether the 4D gauge theory is abelian or non-abelian. 
This form of CS/AS represents our closest analog of the $\Mink_4$ AS analysis developed in Ref.~\cite{Cheung:2016iub}, and also builds on the $\AdS_4^{\Poincare}$ discussion of Ref~\cite{Mishra:2017zan} . We begin with abelian gauge theory for simplicity in this section, and extend to non-abelian gauge theory in the next.

\subsection{Fixed Helicity $\partial \AdS_4$ Correlators}

 In $\Mink_4$ an S-matrix amplitude with an external photon takes the form, 
\begin{align}
\int_{\Mink_4} d^4 X \mathcal{A}_M \mathcal{J}^M \:,
\qquad
\mathcal{A}_M(X) = \epsilon_M^\pm(q) e^{iq \cdot X},
\end{align}
where  $\mathcal{J}$ represents the on-shell current consisting of the rest of the amplitude with amputated photon leg, and $\epsilon_M^{\pm}(q)$ is the polarization vector for $\pm$ helicity, satisfying 
\begin{align}
q^2 = q \cdot \epsilon^\pm = \epsilon^\pm \cdot \epsilon^\pm = 0, \epsilon^\pm \cdot \epsilon^\mp = 1\:.
\label{eq:onShell}
\end{align}
In $\AdS_4$ we compute boundary correlators rather than an S-matrix, 
\begin{align}
\int_{\partial \AdS_4} d^3 x A^\mu (x) 
\left< T\{
J_\mu^{\CFT}(x)\cdots
\}\right>
=
\int_{\AdS_4} d^4 X \mathcal{A}_M(X) \mathcal{J}^M(X)\:, 
\quad \mathcal{A}_\mu(x,z) \xrightarrow[z\rightarrow 0]{} A_\mu (x)\:,
\label{eq:AdS4-corr}
\end{align}
where $\mathcal{A}_M$ satisfies the AdS Maxwell's equations. Given the obvious Weyl invariance of the Maxwell action and the Weyl equivalence of AdS$_4$ to {\it half} of $\Mink_4$, 
\begin{align}
ds^2_{\AdS_4} \:\: \weq \:\: dt^2 - dx^2 - dy^2 - dz^2 \: , \quad z >0\:,
\end{align}
 $\Mink_4$ LSZ wavefunctions for external photons, $\mathcal{A}^\pm_M(X) = \epsilon_M^\pm(q) e^{iq \cdot X} $, are also valid choices for AdS correlators. This corresponds to a $\CFT_3$ source, 
\begin{align}
A^\pm_\mu (x) = \epsilon^\pm_\mu(q) e^{i\hat{q} \cdot x}\:,
\hat{q}\equiv \left( q_0, q_x, q_y\right)\,.
\end{align}
While $A, \mathcal{A}$ are complex, their real and imaginary parts define standard $\partial \AdS/\CFT$ correlators, and we are just considering their complex superposition.  

We choose to work in 4D axial gauge, $\epsilon_z = 0$. It is clear that $A$'s of the above form span all possible sources in $\Mink_3$ with timelike 3-momentum, $\hat{q}$, given that $J$ is conserved (in momentum space, $\hat{q}.J(\hat{q}) = 0$). 
We see that 4D helicity for massless photons matches a 3D ``helicity'' for timelike CFT sources.
The different helicity sources satisfy Chern-Simons-Proca (CSP) equations: 
\begin{align}
2\epsilon^{\mu\nu\rho} \partial_\nu A_\rho = \pm \: m_3 A^\mu\:, \quad m_3 \equiv q_z,
\label{eq:CSP-EOM}
\end{align}
for $\pm$ helicity. Here, $m_3$ is the mass Casimir invariant of $\Mink_3$, that is $m_3^2 = \hat{q}^2 \equiv q_{\mu} q^{\mu}$ for momentum eigenstates, so that $m_3 = q_z$ by Eq.~\eqref{eq:onShell}. 
This has a similar structure to the 3D CSP form of helicity-cut $\Mink_4$ S-matrix amplitudes derived in Ref.~\cite{Cheung:2016iub}, where $m_3$ was the Casimir invariant of a Euclidean AdS$_3$ foliation of (a future light cone in) $\Mink_4$.

\subsection{The  ``holographic soft limit" of $\partial \AdS_4$ correlators}

In $\Mink_4$, it was shown that the conventional (leading) soft photon limit of amplitudes captured by the Weinberg Soft Theorems, was equivalent to the limit $m_3 \rightarrow 0$. Here, we simply translate the analogous definition of ``soft limit'' to the $\AdS_4$ context, as vanishing CSP mass, $m_3 \rightarrow 0$, arriving at the (sourceless) CS equation,  
\begin{align}
\epsilon^{\mu\nu\rho} \partial_\nu A_\rho = 0\:, \quad \partial_\mu A^\mu = 0\:.
\label{eq:CS-From-CSP-EOM}
\end{align}
We also effectively have a Lorentz-gauge fixing condition as can be seen by taking the divergence of the CSP Eq.~\eqref{eq:CSP-EOM} for $m_3 \neq 0$ followed by $m_3 \rightarrow 0$. This gives rise to a ``soft'' $\partial \AdS/\CFT$ correlator, Eq.~\eqref{eq:AdS4-corr}, where
\begin{align}
\mathcal{A}_{\mu}(x,z) = A_\mu(x)\:, \quad \mathcal{A}_z = 0\:.
\end{align}
This follows because $A$ is pure gauge in $\Mink_3$ since $F = 0$ by Eq.~\eqref{eq:CS-From-CSP-EOM}, and therefore this $\mathcal{A}_M$ is pure gauge in $\AdS_4$, hence trivially satisfying 4D Maxwell's equations and $\mathcal{A}_{\mu}(x,z) \xrightarrow [z \rightarrow 0]{} A_{\mu}(x)$.

From the 4D viewpoint, unlike the standard notion of ``soft" in Minkowski spacetime, it is (only)  the holographically emergent direction's $z$-dependence,
 rather than $t$-dependence  (overall energy) which is softened.\footnote{In both Mink$_4$ and AdS$_4$ it is important that the helicity is fixed as we take the soft limit.}  The above 4D pure gauge configurations in the holographic soft limit  are the ``large'' gauge transformations at the root of AS, which we now derive. 
 
\begin{figure}
\centering
\begin{subfigure}[b]{.48\textwidth}
    \centering
    \includegraphics[width=.9\linewidth]{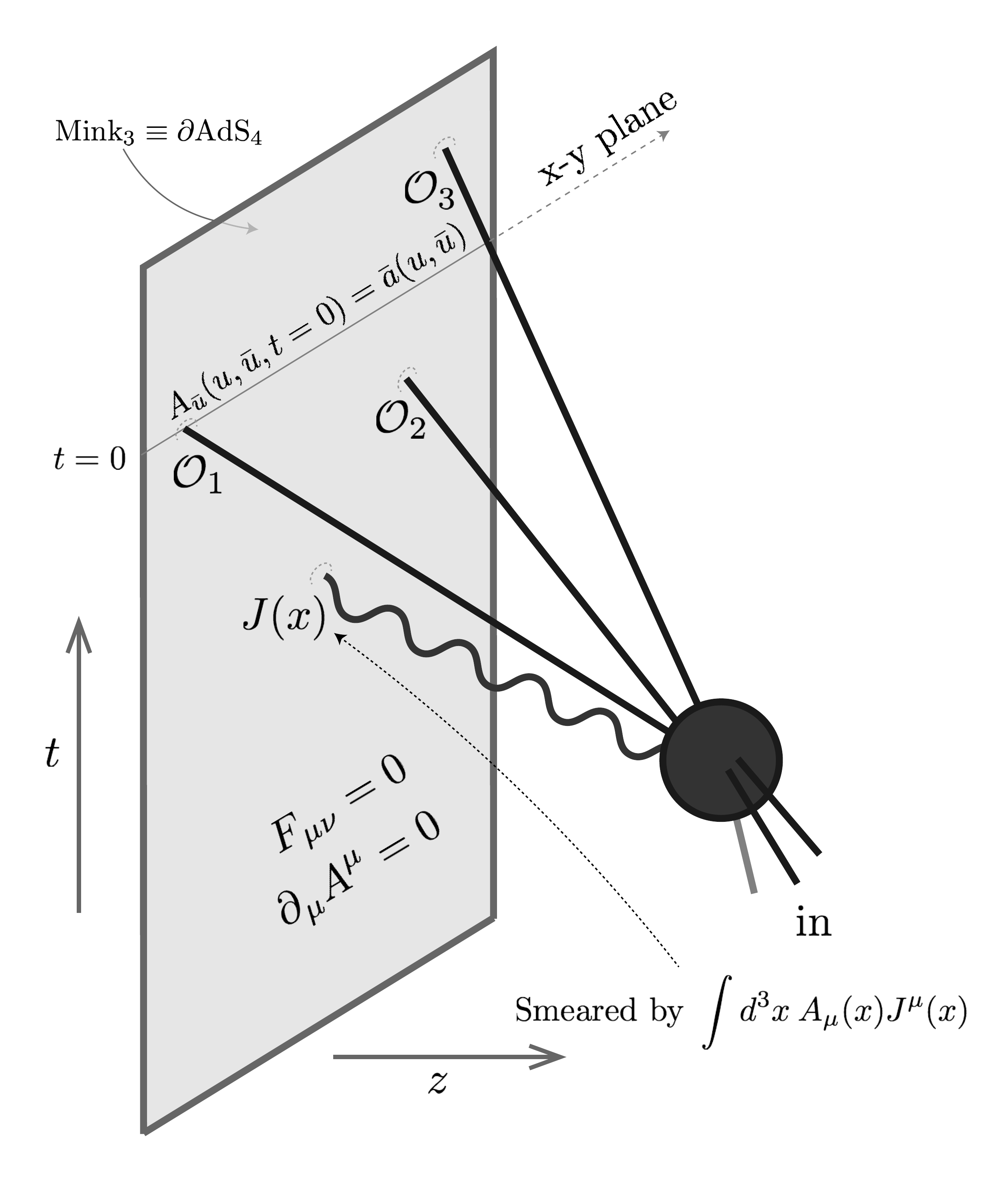}
    \caption{}
    \label{fig:UnequalTimeScatteringAdS-Abelian}
\end{subfigure}\hfill%
\begin{subfigure}[b]{.48\textwidth}
    \centering
    \includegraphics[width=.9\linewidth]{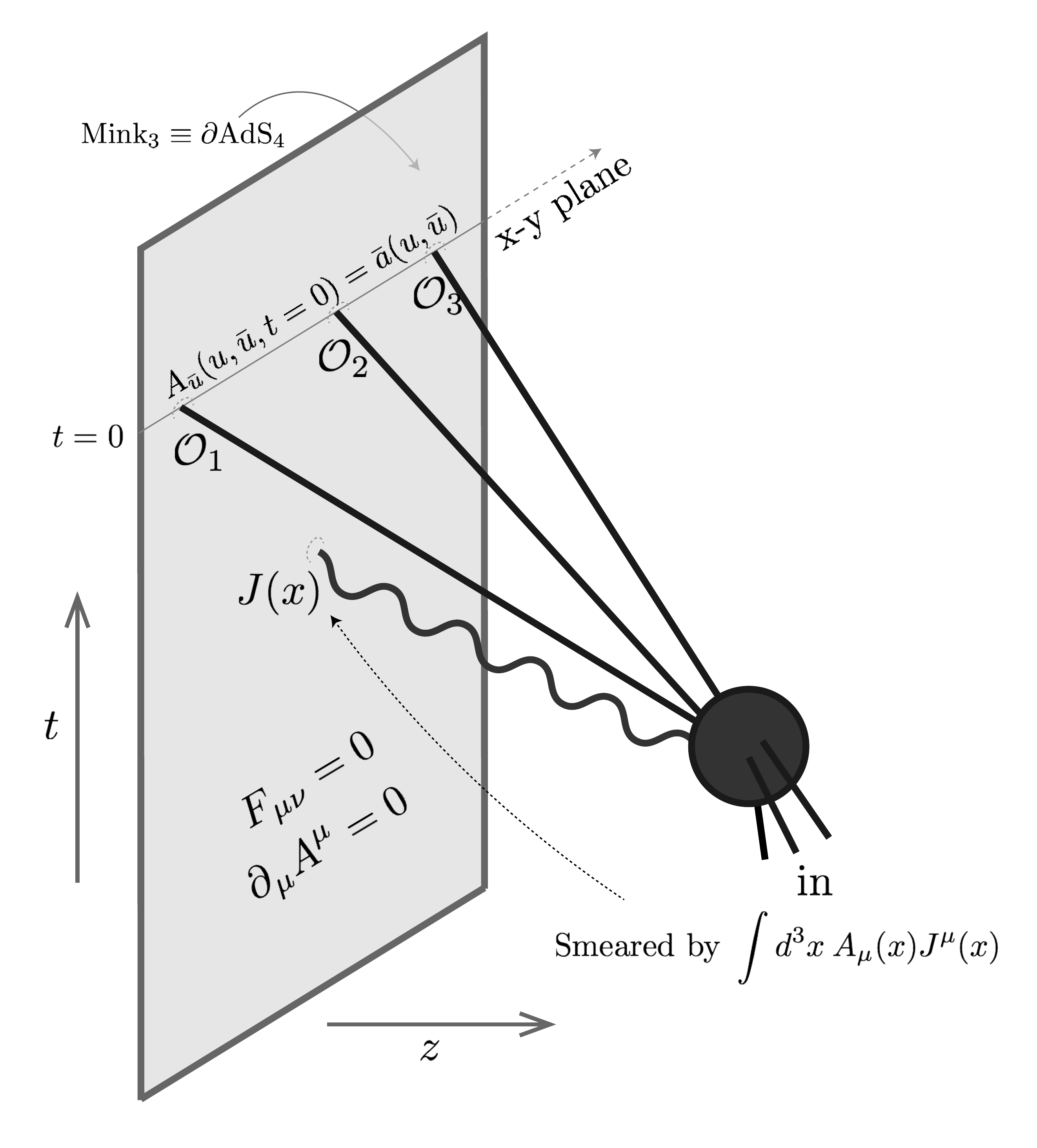}
    \caption{}
    \label{fig:EqualTimeScatteringAdS-Abelian}
\end{subfigure}
\caption{\small{Typical $\partial \AdS_4$ correlators involving 4D photons and matter particles, dual to  $\CFT_3$ correlators of the form Eq.~\eqref{eq:generalAbelianAdSCorrelators} involving the $U(1)$ current and other local operators. (a) corresponds to charged matter lines arriving at general times on the boundary, while (b) corresponds to the special case in which all charged matter arrives at $t=0$.}}
\label{fig:AdS-Abelian-Scattering}
\end{figure}

It is convenient to focus on $\CFT_3$ correlators of the form, 
\begin{align}
\left< 0| T
\left\{
e^{i\int d^3 x \:  A_\mu(x)J^\mu(x)} \, \mathcal{O}_1(x_1)\mathcal \cdots \mathcal{O}_n(x_n)
\right\} | {\rm in}
\right>,
\label{eq:generalAbelianAdSCorrelators}
\end{align}
as depicted in Fig.~\ref{fig:UnequalTimeScatteringAdS-Abelian},
where $A_\mu(x)$ is the source for ``soft" photons,  the ${\cal O}_{\alpha}$ are arbitrary local CFT operators with $U(1)$ charges $Q_{\alpha}$ (including possibly $J^{\mu}$ itself, corresponding to $\partial$AdS correlators for 4D photons which are ``hard" in our sense), and the $\left|\text{in}\right>$ represents a generic initial CFT state.

We write the pure gauge form of $A$ solving the (Lorentz-gauge) CS equations as
\begin{align}
A_\mu(x) = \partial_\mu \lambda(x)\:,\qquad \square_{\Mink_3}\lambda(x) = 0.
\end{align}
We can specify a particular solution in terms of the ``initial'' value ($t=0$), $   \bar{a}(u, {\bar{u}}) \equiv A_{\bar{u}}(u, \bar{u}, t=0) $, first determining

\begin{align}
\lambda( u, \bar{u}, t=0) = \int \frac{d^2 u'}{2\pi} \, \frac{\bar{a}(u',\bar{u}')}{u-u'}\, ,
\label{eq:lambda-zeroTime}
\end{align}
and then uniquely extending to all $t$ once we impose only positive frequencies (absorbing source) in  $\lambda(u, \bar{u}, t)$,
\begin{align}
\lambda(q_u, q_{\bar{u}}, t) &= \lambda(q_u, q_{\bar{u}}, t = 0) \, e^{-2i\sqrt{q_u q_{\bar{u}}}\: t}.
\label{eq:lambda-nonZeroTime}
\end{align}

By the CFT current Ward identity, 
\begin{equation}
\partial_{\mu} J^{\mu} = -\sum_{\alpha} Q_{\alpha} \delta^3(x - x_{\alpha}),
\label{eq:CFT-Ward-Identity}
\end{equation}
 we find
 \begin{align}
i\int d^3 x A_\mu(x) J^\mu (x) 
= 
i\sum_\alpha Q_{\alpha} \lambda(x_{\alpha}) 
\label{eq:A-dot-J}
\end{align} 

\subsection{2D Holomorphic Abelian WZW Current from Holographic Soft Limit}

Let us focus first on the special case that all the ${\cal O}_{\alpha}$ are simultaneous,  $t_{\alpha} = 0$, as depicted in Fig.~\ref{fig:EqualTimeScatteringAdS-Abelian}, so that by Eqs.~\eqref{eq:A-dot-J}, \eqref{eq:lambda-zeroTime}, 
\begin{align}
i\int d^3 x A_\mu(x) J^\mu (x) 
= 
-i\int \frac{d^2 u}{2\pi} \, \bar{a}(u, \bar{u}) \sum_\alpha \frac{Q_\alpha}{u - u_\alpha} \: .
\end{align}
Thinking of $\bar{a}(u, \bar{u})$ as a source defining a 2D current $j \equiv 2\pi i \,\delta/\delta \bar{a}(u, \bar{u})$, we arrive at a 2D holomorphic form for $j$,
\begin{align}
\left< 0| 
j(u, \bar{u})
\mathcal{O}_1(x_1)\cdots \mathcal{O}_n(x_n)
| {\rm in}  \right>
=
\sum_\alpha
\frac{Q_\alpha}{u - u_\alpha}
\left< 0| 
\mathcal{O}_1(x_1)\cdots \mathcal{O}_n(x_n)
| {\rm in} \right>.
\label{eq:ope-Abelian}
\end{align}
The simple pole structure of $j$ is clearly very similar to that observed in soft limits of the Mink$_4$ S-matrix.
We can straightforwardly obtain multiple-$j$ correlators since the source is simply exponentiated, but there is 
 no central extension singularity in $j j $ correlators as they coincide, for reasons further discussed in the next section.

In the general case of non-simultaneous $t_{\alpha}$ (Fig.~\ref{fig:UnequalTimeScatteringAdS-Abelian}),  Eq.~\eqref{eq:A-dot-J} gives  
a 2D current defined by source $\bar{a}$,  
\begin{align}
j(u, \bar{u})  
=
-2\pi\,\sum_\alpha Q_\alpha \frac{\delta \lambda(\bar{a}, x_\alpha)}{\delta \bar{a}(u, \bar{u})},
\label{eq:j-non-simultaneous}
\end{align} 
but this is no longer holomorphic, reminiscent of the case of massive charges in the Mink$_4$ S-matrix.
We explore this non-holomorphic structure more closely in Section~\ref{sec:memory} in the context of the memory effect.

\section{Non-abelian Generalization of Holographic Soft Limit and AS}
\label{sec:MirrorDual-from-SoftLimit-nonAbelian}

There is a natural generalization of ``soft"  to (tree-level) {\it non}-Abelian AdS$_4$ gauge theory. Generalizing Eq.~\eqref{eq:AdS4-corr}, we consider a 4D ``soft" field ${\cal A}^{a}_{M}$ which is a complex solution to the 4D Yang-Mills equations, coupled to a 4D gauge current ${\cal J}^a_M$ representing other charged matter and ``hard'' gluons. The boundary limit ${\cal A}^{a}_{\mu} \xrightarrow[]{z \rightarrow 0} A^a_{\mu}$ of such a complex solution simply corresponds to a complex source $A_{\mu}$ for $J^{\CFT}_{\mu}$ and its associated CFT correlators. 

When there are multiple ``soft'' gluons, we must generalize the fixing of helicity of ``soft'' photons in the Abelian case in a manner that is compatible with Yang-Mills self-couplings. This is given by requiring the {\it complex} ${\cal A}^{a}_{M}$ to be self-dual (or alternatively, anti self-dual):
\begin{equation}
\frac12 \epsilon^{\mu \nu \rho} {\cal F}^a_{\mu \nu}(x,z) = i  {\cal F}^{\rho z~a}(x,z)
\quad  \underset{\text{axial gauge}}{=}  
\quad
i \partial_z {\cal A}^{a~\rho}(x,w),  
\end{equation}
where ${\cal F}$ is the full non-abelian 4D field strength. 
This is closely analogous to what is seen in 4D Minkowski spacetime, where the non-abelian soft ``branches'' attached to a hard scattering process are self-dual when all its external soft gluons have positive helicity~\cite{Cheung:2016iub}.

In axial-gauge, the holographic soft limit is again that in which ${\cal A}^{a}_{\rho}$ is $z$-independent. Self-duality then implies the vanishing of all of ${\cal F}$, so that ${\cal A}$ is pure-gauge. The CFT source is simply  given by $A_{\mu}^a \equiv {\cal A}^a_{\mu}(x,z \rightarrow 0) = {\cal A}^a_{\mu}(x)$, so that it satisfies a (sourceless) non-Abelian CS equation,
\begin{align}
\epsilon^{\mu\nu\rho} F_{\nu \rho}^a(x) = 0,
\label{eq:CS-EOM-NonAbelian}
\end{align}
again closely analogous to the Mink$_4$ analysis. 
More precisely, there will also be an effective 3D gauge-fixing condition that results from the approach to the soft limit, but it will be more complicated than the simple 3D Lorentz gauge of the Abelian case, Eq.~\eqref{eq:CS-From-CSP-EOM}. As for the Abelian case, this condition will not be relevant for the special case of {\it equal-time} correlators of CFT local operators, to which we now turn.

\subsection{2D Holomorphic Non-abelian WZW Current from Holographic Soft Limit}

The vanishing of the non-Abelian field strength of the source in the soft limit has the solution,
\begin{align}
iA_\mu(x) = e^{-i\lambda(x)}\partial_\mu e^{i\lambda(x)}\:, \quad \lambda \equiv \lambda^a t^a, \: A_\mu \equiv A_\mu^a t^a\:,
\label{eq:pure-gauge-non-abelian}
\end{align}
where the $\lambda^a(x)$ are {\it complex} gauge transformation fields, reflecting the complex nature of $A_{\mu}^a$ (necessary for Lorentzian self-dual gauge fields). Starting from the general correlator, 

\begin{align}
\left< T
\left\{
e^{i\int d^3 x \:  A^a_\mu(x)J^{\mu a}(x)} \, \mathcal{O}_1(x_1)\mathcal \cdots \mathcal{O}_n(x_n)
\right\}
\right>\:,
\end{align}
we will again consider $\bar{a}^a(u, \bar{u}) \equiv A^a_{\bar{u}}(u, \bar{u}, t=0)$ as the independent variables behind our soft source $A_{\mu}(x)$, and define a 2D current 
\begin{align}
j^a(u,\bar{u}) \equiv 2\pi i \,\frac{\delta}{\delta \bar{a}^a(u,\bar{u})}\:.
\end{align}

For single $j$ correlators with equal-time ``hard" operators, $t_{\alpha} = 0$, the non-Abelian structure is clearly irrelevant, and we arrive at the analog of Eq.~\eqref{eq:ope-Abelian} again,

\begin{align}
\left< 0| 
j^a(u, \bar{u})
\mathcal{O}_1(x_1)\cdots \mathcal{O}_n(x_n)
|{\rm in} \right>
=
\sum_\alpha
\frac{T_{(\alpha)}^a}{u - u_\alpha}
\left< 0| 
\mathcal{O}_1(x_1)\cdots \mathcal{O}_n(x_n)
|{\rm in} \right> \: .
\label{eq:ope-NonAbelian}
\end{align}

\begin{figure}
\centering
\includegraphics[width=.5\linewidth]{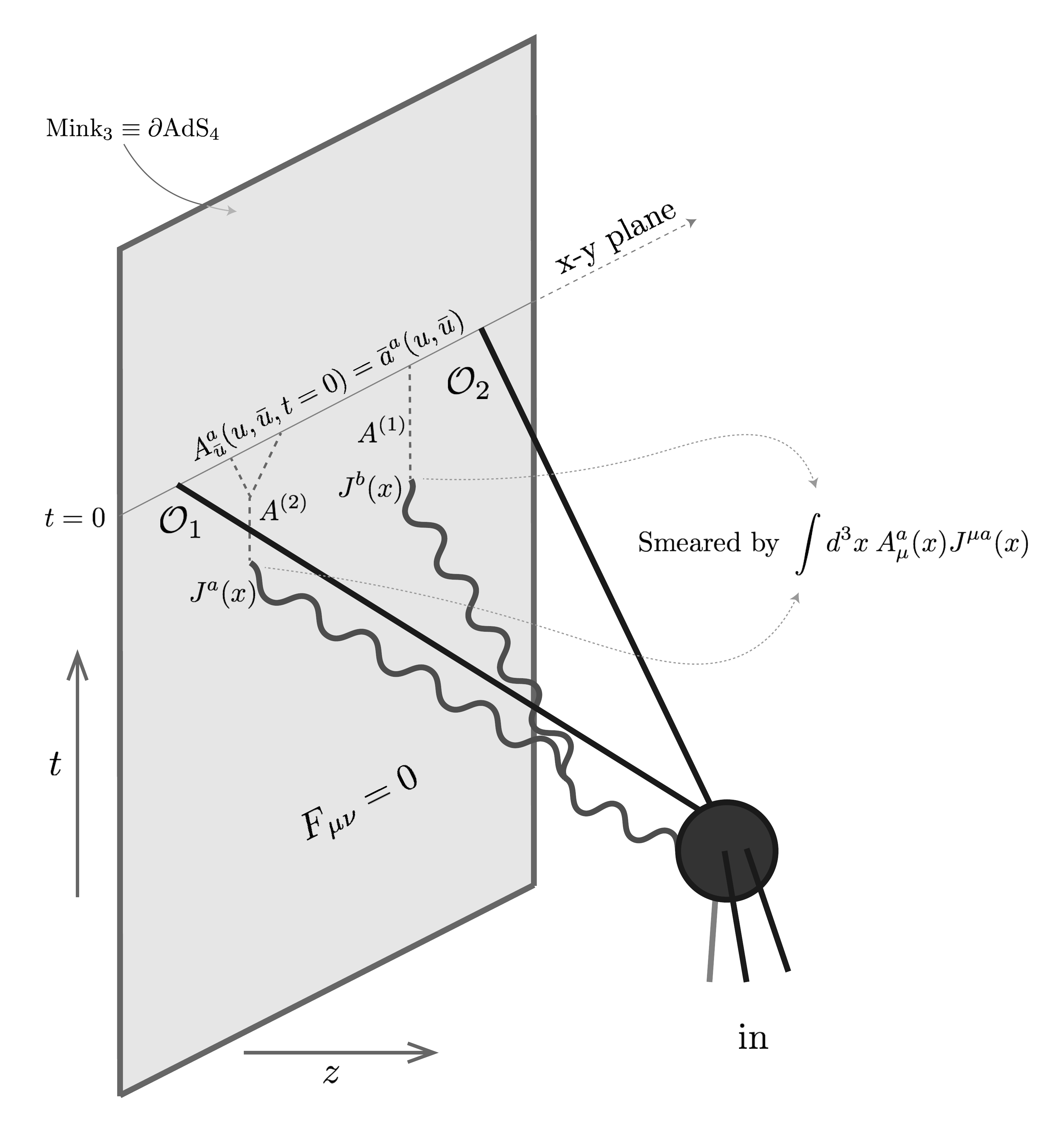}
\caption{\small{A typical $\partial \AdS_4$ correlator for non-abelian AdS gauge theory, with all hard matter arriving at $t=0$. Note that there are both non-abelian bulk interactions and non-abelian corrections to the ``softened" source in terms of the independent variables $\bar{a}^a$. The leading source term $A^{(1)}$ is similar in form to the abelian case, while the next non-abelian correction $A^{(2)}$ is given by the last two terms in Eq.~\eqref{eq:A-to-second-order}.}}
\label{fig:ScatteringAdS-nonAbelian}
\end{figure}

Next we probe correlators $\langle j^a(u, \bar{u})  j^b (u', \bar{u}') ... \rangle$, to search for a non-abelian contribution to the $j j'$ 2D ``OPE''.
This requires us to work to order $\bar{a}^2$.  At first order in $\bar{a}$, we obviously have 
\begin{align}
\lambda^{(1)\, a} ( u, \bar{u}, t=0)
=
\int \frac{d^2 u'}{2\pi} \: \frac{\bar{a}^a (u', \bar{u}')}{u-u'}\:,
\end{align}
as in the Abelian case.  To second order, by Eq.~\eqref{eq:pure-gauge-non-abelian},
\begin{align}
A_\mu^a(x) \approx 
\partial_\mu \lambda^{(1)\, a}(x)
- \frac{1}{2}f^{abc}\lambda^{(1)\, b}(x) \partial_\mu \lambda^{(1)\, c}(x)
+ \partial_\mu \lambda^{(2)\, a}(x)\: .
\label{eq:A-to-second-order}
\end{align}
We can use the $\bar{u}$ component of this to solve for $\lambda^{(2)}(t=0)$, 
\begin{align}
\partial_{\bar{u}}\lambda^{(2)\, a}(u, \bar{u}, t=0)
&=
\frac{1}{2} f^{abc}\lambda^{(1)\, b} (u, \bar{u}, t=0)\: \bar{a}^c(u, \bar{u}) \: , 
\end{align}
from which we derive
\begin{align}
\lambda^{(2)\, a}(u, \bar{u}, t=0)
&=
\frac{1}{2} f^{abc}
\int \frac{d^2 u'}{2\pi} \: \int \frac{d^2 u''}{2\pi} \:
\frac{\bar{a}^b(u', \bar{u}') \, \bar{a}^c(u'',\bar{u}'')}{(u-u'')(u'' - u')}
\: . 
\label{eq:lambda2}
\end{align}
In this way we see two types of non-abelian corrections enter into the typical $\partial \AdS_4/\CFT_3$ correlator compared to the abelian case, as depicted in Fig.~\ref{fig:ScatteringAdS-nonAbelian}. Of course there are non-abelian interactions in the 4D bulk, but we also have non-abelian corrections to the CFT ``softened" source $A^a_{\mu}$ when expressed in terms of the independent variables $\bar{a}^a$.

We see that Eq.~\eqref{eq:lambda2} can give rise to a non-trivial ``OPE'' divergence for coinciding $j$'s, so we drop $\lambda^{(1)}$ contributions to focus on that of $\lambda^{(2)}$:
\begin{align}
\int d^3 x \; A_\mu^a(x) J^{\mu\, a}(x) 
&\supset 
\int d^3 x \; \partial_\mu \lambda^{(2)\,a}(x) J^{\mu\, a}(x)
\nonumber \\
&= - \int d^3 x \; \lambda^{(2)\,a}(x) \partial_\mu J^{\mu\, a}(x) 
= \sum_\alpha \lambda^{(2)\,a}(x_\alpha) T^a_{(\alpha)}\:.
\end{align}

Specializing to the simultaneous limit, $t_{\alpha} = 0$,
\begin{align}
\int d^3 x \; A_\mu^a(x) J^{\mu\, a}(x)  
&\supset  
\: \sum_\alpha \lambda^{(2)\,a}(u_\alpha, \bar{u}_\alpha, t_\alpha = 0) \, T^a_{(\alpha)}
\nonumber \\
&=
\frac{1}{2} f^{abc}
\int \frac{d^2 u}{2\pi} \: \int \frac{d^2 u'}{2\pi} \:
\frac{\bar{a}^b(u, \bar{u}) \, \bar{a}^c(u',\bar{u}')}{(u_\alpha-u')(u' - u)}
\, T^a_{(\alpha)}\: .
\end{align}
We thereby derive,
\begin{align}
& 
\left< 0|T\left\{ 
j^a(u, \bar{u})
j^b(u', \bar{u}')
\mathcal{O}_1(x_1)\cdots \mathcal{O}_n(x_n)
\right\}
| {\rm in} 
\right>
\nonumber \\
&  \qquad \qquad \supset \:\:
\frac{1}{2} f^{abc}
\sum_\alpha
T^c_{(\alpha)}
\left\{
\frac{1}{(u_\alpha - u)(u- u')}
-
\frac{1}{(u_\alpha - u')(u'- u)}
\right\}
\left< 0| 
\mathcal{O}_1(x_1)\cdots \mathcal{O}_n(x_n)
| {\rm in}  \right>
\nonumber \\
& \qquad \qquad \underset{u' \rightarrow u}{\sim}
\frac{f^{abc}}{u' - u}
\sum_\alpha \frac{T_{(\alpha)}^c}{u-u_\alpha} 
\left< 0| 
\mathcal{O}_1(x_1)\cdots \mathcal{O}_n(x_n)
| {\rm in}  \right>
\nonumber \\
& \qquad \qquad =  \:\:
\frac{f^{abc}}{u' - u}
\left< 0|
T\left\{j^c(u,\bar{u}) \mathcal{O}_1(x_1)\cdots \mathcal{O}_n(x_n)\right\}
| {\rm in}  \right>\: .
\end{align}

In this sense, we have arrived at the Euclidean 2D KM ``OPE'', 
\begin{align}
j^a(u,\bar{u})\;j^b(u',\bar{u}') 
\:\: \underset{u\rightarrow u'}{\sim} \:\:
\frac{f^{abc}}{u-u'}\:j^c(u, \bar{u}),
\end{align}
but unlike the canonical Eq.~\eqref{eq:jj-OPE-Abelian} we see that we have vanishing central extension here!  This absence of a central extension in AS from soft limits matches what is seen in 4D Minkowski spacetime. But as pointed out in Ref.~\cite{Cheung:2016iub}, it is closer to the truth to say that we have {\it infinite} central extension, as we review below.

\subsection{Holographic Soft Limit as Portal from Standard to Alternate Theory}

The structure of correlators of $j$ we see in the holographic soft limit with Dirichlet boundary condition precisely matches that found  in Ref.~\cite{Mishra:2017zan} for alternate b.c in the $\kappa \rightarrow \infty$ limit, as shown there by simple $\kappa$-counting diagrammatic arguments. Here, we just give a heuristic argument for why this is so, based on the path integral for {\it dynamical} CS coupled to the CFT (dual to alternate boundary condition),
\begin{align}
\int \mathcal{D} A_\mu
\exp\left\{
i\int d^3 x \: \frac{\kappa}{4\pi} \,\epsilon^{\mu\nu\rho}\,\text{Tr}\,
\left(
A_\mu \partial_\nu A_\rho
+\frac{2}{3} \,A_\mu A_\nu A_\rho
\right)
+ A_\mu^a\, J^{\mu\,a}_{\CFT_3}
\right\}
\:.
\end{align}
We see that as the CS level $\kappa \rightarrow \infty$,  there is a wild phase in the path integral, forcing the $\kappa$-dependent part of the action to be extremized,
yielding Eq.~\eqref{eq:CS-EOM-NonAbelian}, derived here via the ``soft" limit. With the $t=0$ condition on the path integral, $A_{\bar{u}}^a(t=0) \equiv \bar{a}^a$ (and gauge-fixing), this leads to a specific $A_{\mu}^a(x)$. In this way, the alternate boundary condition becomes effectively Dirichlet boundary condition as $\kappa \rightarrow \infty$, in particular matching the holographic soft limit. The one ``flaw" with this argument is that the $\kappa \rightarrow \infty$ limit for dynamical $A$  is ill-defined for $j j'$ correlators, precisely because of the central term in Eq.~\eqref{eq:jj-OPE-Abelian}. As pointed out in Ref.~\cite{Mishra:2017zan}, this is avoided by only considering connected correlators of the CS fields {\it with} the CFT, since the central term arises from connected correlators of CS with only itself. From the Dirichlet boundary condition viewpoint, this restriction is automatic since we are always considering soft dressing  of ``hard'' CFT correlators. With this restriction, the central extension of KM is 
absent, as if it vanished, when in fact it is infinite as $\kappa \rightarrow \infty$.

The seeds of alternate boundary condition correlators  are contained in 
the Dirichlet boundary condition AdS$_4$ (pure CFT$_3$)  correlators via their holographic soft limits. One can then unitarize these leading-in-$\kappa$   correlators 
by going to finite large $\kappa < \infty$, and including the simple pure-CS correlators, which contain the central extension. In this nuanced sense, AS from soft limits are a remnant of the alternate b.c theory, dual to the CS-gauged CFT$_3$. 

\section{CS Memory Effects and the Holographic Soft Limit}
\label{sec:memory}

Finally, we point out that AdS$_4$ gauge theory exhibits an analog of the electromagnetic ``memory'' phenomenon of Mink$_4$~\cite{Susskind:2015hpa,Pasterski:2015zua,Strominger:2015bla,Strominger:2017zoo}, closely connected to  AS structure. The memory effect 
compares the parallel transport between two test charges far from a scattering process, long before and after the scattering event, more precisely given by a Wilson loop consisting of spatial transport between the two charges at early and late times, and temporal transport  between those times. 
We focus on the abelian case. 

\subsection{Alternate boundary conditions and electric memory}

\begin{figure}
\centering
\includegraphics[width=.7\linewidth]{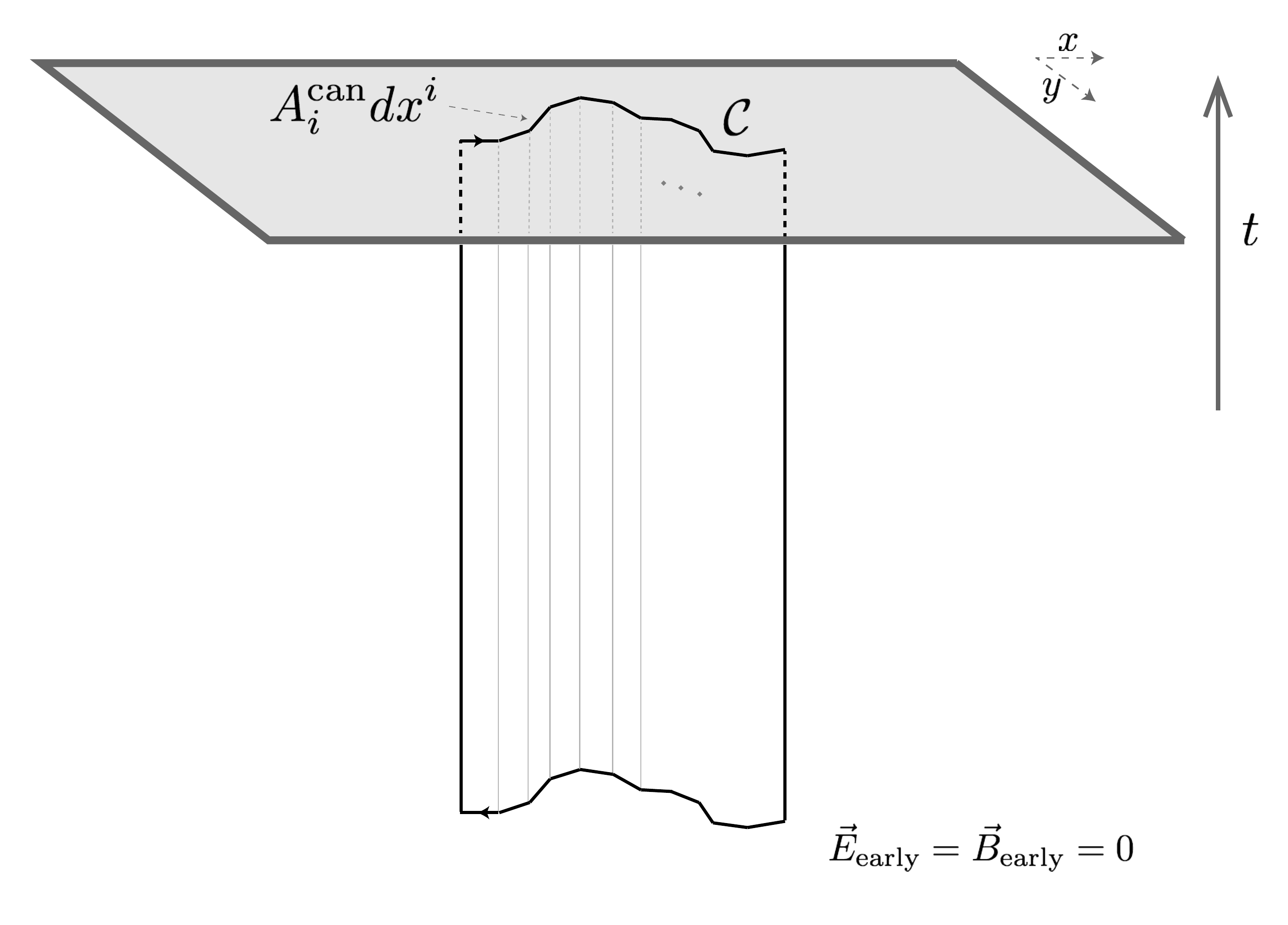}
\caption{\small{A general CS memory Wilson loop in Mink$_3$, comparing parallel transport along the spatial curve 
${\cal C}$ at early and late times, where for simplicity the early state has vanishing gauge field strength. It can be viewed as composed of many narrow memory Wilson loops, with shared timelike lines canceling due to their opposing orientations. In terms of the canonical CS fields, effectively in temporal gauge, this general Wilson loop is therefore given by just the late Wilson line along ${\cal C}$. (See Eq.~\eqref{eq:CS-as-Wilson-loop})}}
\label{fig:CS-as-memory}
\end{figure}

We begin with alternate boundary condition, in its dual formulation as $U(1)$ CS + CFT$_3$. Canonically, the CS fields are  $A_{\bar{u}}, A_{u}$, effectively in temporal gauge $A_0 = 0$ after deriving the Gauss Law constraint. For simplicity focussing on vanishing electromagnetic field strengths at early times (hence only neutral particles in the initial state), we can choose the further gauge condition $A_u(t = -\infty), A_{\bar{u}}(t= - \infty) =0$. 
We see that our canonical (can) fields therefore precisely define ``memory'' Wilson loops in more general (gen) gauges, 
\begin{align}
A_i^{\text{can}}(u, \bar{u}, t=0)dx^i 
&=
A_i^{\text{gen}}(u, \bar{u}, t=0)dx^i 
+ \int_0^{-\infty} dt' A_0^{\text{gen}}(u+du, \bar{u}+d\bar{u}, t')
\nonumber \\
& -A_i^{\text{gen}}(u, \bar{u}, t=-\infty)dx^i 
+ \int_{-\infty}^0 dt' A_0^{\text{gen}}(u, \bar{u}, t')\:, 
~~ {\rm where ~} i \equiv u,\bar{u}.
\label{eq:CS-as-Wilson-loop}
\end{align}
The four terms on the right define four sides of a narrow gauge-invariant ``memory'' Wilson loop, from $u$ to $u+du$ at time $t=0$, to time $- \infty$ at $u+du$, back from $u+du$ to $u$ at time $- \infty$, and then from time $- \infty$ to $t=0$ at $u$. Similarly, a Wilson line of $A^{\text{can}}$ along a finite spatial curve ${\cal C}$ in the $x-y$ plane at time $t=0$ is equivalent to 
a more general memory Wilson {\it loop} in a general gauge, completing the curve with time-like lines to $t = - \infty$ and spatial Wilson line reversing ${\cal C}$ at time $- \infty$. This is depicted in Fig.~\ref{fig:CS-as-memory}.
Because of the Gauss Law constraint, the precise choice of  ${\cal C}$ does not matter as along as one does not cross 3D charges in deforming the curve. 

In the above sense, arbitrary CS gauge theories describe the dynamics of memory effects in 3D. But when the CS charged matter is a CFT$_3$ with AdS$_4$ dual, the memory effects ``lift'' to 4D. 
The 3D memory Wilson loop above is now seen as a 4D memory Wilson loop at (or near) $\partial \AdS_4$, $z=0$, far from a bulk scattering. This is similar to the $\Mink_4$ memory Wilson loops at large distance from a scattering process~\cite{Susskind:2015hpa,Pasterski:2015zua,Strominger:2015bla}. In the alternate boundary condition AdS case, the CS Gauss Law gives a general relationship between the canonical memory fields $A_{u}^{\text{can}}, A_{\bar{u}}^{\text{can}}$. As noted in subsection~\ref{subsec:GaussLawConstraint},  this relationship effectively determines the CS quantum state completely in terms of the matter CFT state, say as a wavefunctional in $A_{\bar{u}}^{\text{can}}$ in coherent state basis. Both $A_{u}^{\text{can}}$ and  $A_{\bar{u}}^{\text{can}}$ are determined as operators acting on this state. 
That is, Gauss' Law completely determines the memory effect at the quantum level.  As we saw in subsections~\ref{subsec:2DWZWCurrent} and~\ref{subsec:generalCFTstates} Gauss' Law is essentially equivalent to the KM structure. Thus, at the most fundamental level, the memory effect is the physical face of the AS structure.

\subsection{Dirichlet boundary conditions and magnetic memory}

Let us switch to Dirichlet boundary condition, in which case the boundary-localized memory Wilson loop vanishes (dual to the absence of CS fields, given just the isolated CFT$_3$). But we saw in Section~\ref{sec:EMduality-from-MirrorDuality} that in magnetic dual variables $\widetilde{\mathcal{A}}$ the boundary condition becomes effectively Neumann. This allows us to consider non-vanishing magnetic memories, given by  't Hooft loops (Wilson loops in $\widetilde{\mathcal{A}}(z\rightarrow 0)$). We will see that 
this can be non-trivial even in processes involving only standard electric charges but no magnetic charges. These effects are analogous
to (the electric-magnetic dual of) the  magnetic memory effects in $\Mink_4$ discussed in Ref. \cite{Strominger:2015bla}.

There is an important but subtle contrast with the previous subsection. From the holographic viewpoint of the magnetic dual description, there is a 3D  $\widetilde{A}$ which is the  mirror version of $A$ above. Naively, this $\widetilde{A}$ translates via AdS/CFT into $\widetilde{\cal A}(z=0)$ in the 4D description. But formally $\widetilde{A}$ has a CS level $\widetilde{\kappa} = 0$, so that rather than being a CS field it reduces to a simple Lagrange multiplier for $\widetilde{J}$,  which  translates via AdS/CFT  to  the Lagrange multiplier enforcing the Neumann boundary conditions in AdS. Thus, once we are considering the 4D magnetic dual description with Neumann boundary conditions, {\it this} $\widetilde{\cal A}(z=0)$ has already been integrated out of the theory. Instead, in this subsection we are considering the distinct Neumann bulk field $\widetilde{\mathcal{A}}(z)$ in the limit $z \rightarrow 0$. Unlike $A$ (or $\widetilde{A}$), $\widetilde{\mathcal{A}}_u(z\rightarrow 0)$ and $\widetilde{\mathcal{A}}_{\bar{u}}(z\rightarrow 0)$ are {\it not} canonically conjugate, and are not constrained by a (mirror) Gauss Law constraint.

We begin with the standard $\AdS/\CFT$ identification of holographic charge density, 
\begin{align}
J_0^{\CFT} \equiv \frac{1}{g^2}{\cal F}_{0z}(z \rightarrow 0) = \frac{1}{g^2}\widetilde{\cal F}_{xy}(z \rightarrow 0) =
\frac{-2i}{g^2}
\left(
\partial_u \widetilde{\mathcal{A}}_{\bar{u}}(z\rightarrow 0)-\partial_{\bar{u}}\widetilde{\mathcal{A}}_u(z\rightarrow 0)
\right)\, .
\label{eq:da-minus-daBar}
\end{align}
Note that this relates the magnetic $\widetilde{\mathcal{A}}(z\rightarrow 0)$ gauge field with the original electric CFT$_3$ current.  
For given $J$, this is a general constraint on the memories measured by the (temporal gauge) $\widetilde{\mathcal{A}}_u(z\rightarrow 0), \widetilde{\mathcal{A}}_{\bar{u}}(z\rightarrow 0)$.

\begin{figure}
\centering
\includegraphics[width=.5\linewidth]{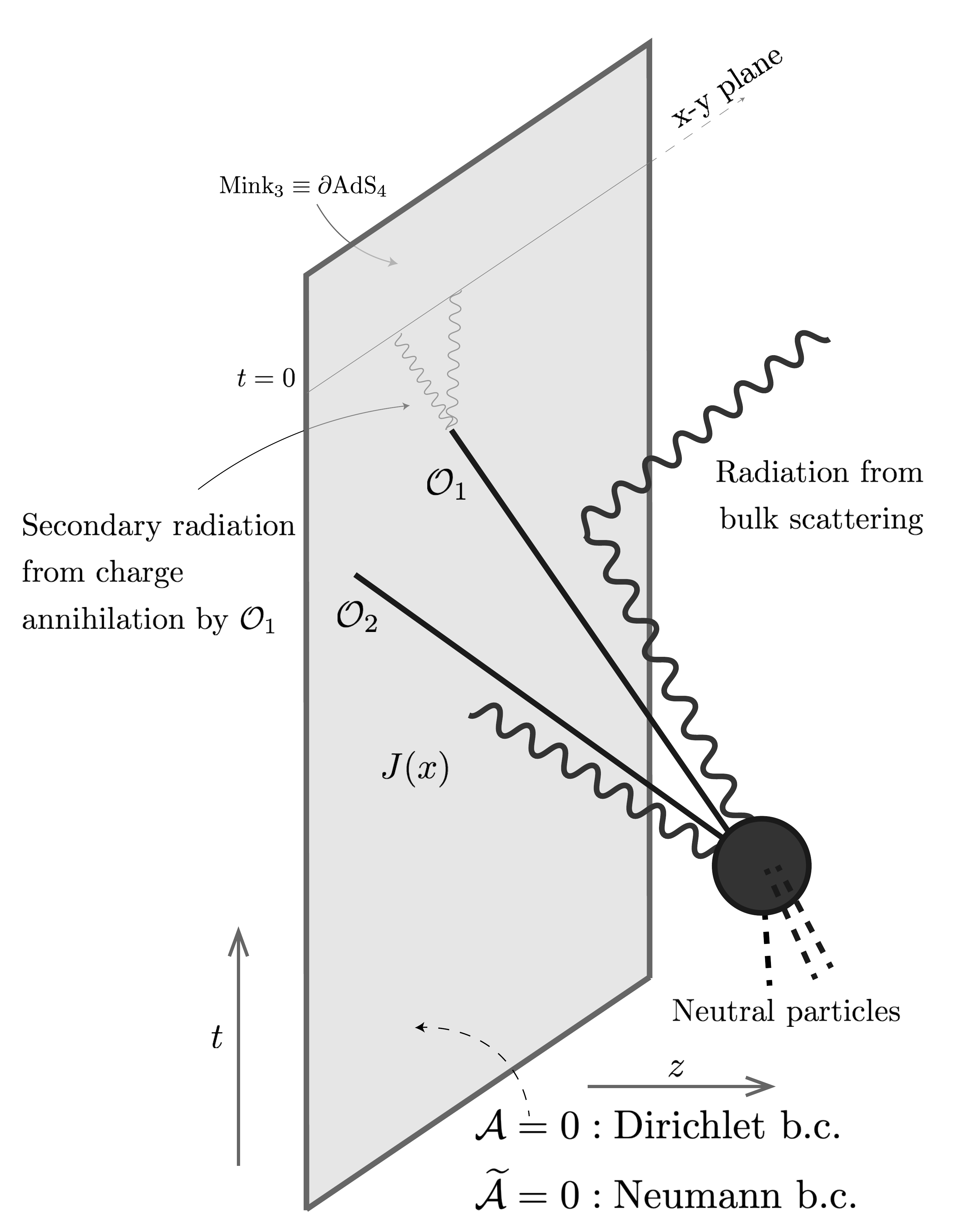}
\caption{\small{A $\partial \AdS_4$ correlator for radiation and charged matter created by a distant bulk scattering, initiated from an electromagnetically neutral state. We focus on a 't Hooft line at $t=0$ in temporal gauge, corresponding to a magnetic memory loop, allowed by the standard boundary conditions. It receives contributions from the secondary radiation emitted by charged matter annihilated at the boundary by local operators. Radiation from the bulk scattering is either absorbed by the CFT current $J$ or reflected by the boundary, and therefore does not contribute to the late-time 't Hooft line.}}
\label{fig:MemoryInAdS}
\end{figure}

In special circumstances, analogous to the set-up in Mink$_4$, we can make a stronger statement.
  We will assume that our initial state has vanishing field strengths, involving a non-trivial scattering of neutral particles deep in the bulk of AdS$_4$, and results in production of 4D electromagnetic radiation and electrically (not magnetically) charged particles. We take the charges to be massless so that we can continue to treat AdS$_4$ as effectively Mink$_4/2$ by Weyl invariance, and take local CFT operators ${\cal O}_{\alpha}(x_{\alpha})$  to annihilate the charges on $\partial \AdS_4$ at $t_{\alpha} < 0$, before the memory measurement at $t=0$.  More generally, we take the radiation and particles to arrive at $\partial \AdS_4$ earlier than $t=0$, and either be reflected away into the bulk or absorbed by boundary/CFT operators. Therefore, radiation from the bulk scattering does not contribute to the boundary $\widetilde{\mathcal{A}}(z\rightarrow 0)$ gauge fields at $t=0$. This set-up is depicted in Fig.~\ref{fig:MemoryInAdS}.

But further radiation can result when the charged particles are absorbed by ${\cal O}_{\alpha}$ on $\partial \AdS_4$, effectively ``annihilating'' with their images in the Mink$_4$ covering space of Mink$_4/2 \sim \AdS_4$. This secondary radiation from $z \sim 0$ can spread until $t=0$ and contribute to the boundary fields $\widetilde{\mathcal{A}}(z\rightarrow 0)$ then. In temporal gauge, the transverse radiation satisfies $\partial_x \widetilde{\mathcal{A}}_x(z\rightarrow 0) + \partial_y\widetilde{\mathcal{A}}_y(z\rightarrow 0) + \partial_z \widetilde{\mathcal{A}}_z(z\rightarrow 0)= 0$ as usual. Since the secondary radiation travels in the $x-y$ directions but remains at $z \sim 0$ in order to contribute to the memory measurement there, the $z$-momentum is subdominant, and we have 
\begin{align}
\partial_x \widetilde{\mathcal{A}}_x(z\rightarrow 0) + \partial_y \widetilde{\mathcal{A}}_y(z\rightarrow 0) 
\equiv \partial_u \widetilde{\mathcal{A}}_{\bar{u}}(u, \bar{u}, z \rightarrow 0, t=0) +  \partial_{\bar{u}} \widetilde{\mathcal{A}}_{u}(u, \bar{u}, z \rightarrow 0, t=0) 
\approx 0. 
\label{eq:da-plus-daBar}
\end{align}
 We can then solve the simultaneous equations, Eqs.~\eqref{eq:da-minus-daBar}, \eqref{eq:da-plus-daBar}, for the memory fields, 
\begin{align}
\widetilde{\mathcal{A}}_u(u,\bar{u}, z \rightarrow 0, t=0) &= -\frac{ig^2}{4} \int \frac{d^2 u'}{2\pi} \frac{J_0(u', \bar{u}', t=0)}{u' - u}
\nonumber \\
\widetilde{\mathcal{A}}_{\bar{u}}(u,\bar{u}, z \rightarrow 0, t=0) &= \frac{ig^2}{4} \int \frac{d^2 u'}{2\pi} \frac{J_0(u', \bar{u}', t=0)}{\bar{u} - \bar{u}'}.
\label{eq:a-aBar-Solution}
\end{align}

We now show that the above memory effect precisely matches the holographic soft limit we derived in Section~\ref{sec:MirrorDual-from-SoftLimit-abelian}. 
First we note that the secondary radiation satisfies the Maxwell equations, 
\begin{align}
0 &= \partial_i B_i + \partial_z B_z \approx \partial_i B_i
\nonumber \\
0 &= \partial_0 B_i + \epsilon_{ij}\partial_j E_z - \epsilon_{ij} \partial_z E_j
\approx \partial_0 B_i + \epsilon_{ij}\partial_j E_z,  ~ ~ {\rm where} ~ i \equiv x,y,
\end{align}
and where again the $z$-momentum is subdominant so that we drop the $\partial_z$ terms. Since we are near the boundary, we can translate $B_i \rightarrow g^2 \epsilon_{ij} J_j$ and $E_z \rightarrow g^2 J_0$, so that the above relations become 
\begin{align}
\epsilon^{\mu\nu\rho}\partial_\nu J_\rho \approx 0\,.
\end{align}
Therefore $J_{\mu} \approx \partial_{\mu} \Phi$ is a total gradient. The current Ward identity, Eq.~\eqref{eq:CFT-Ward-Identity}, then reads
\begin{align}
\partial_\mu \partial^\mu \Phi = - \sum_\alpha Q_\alpha \delta^3(x-x_\alpha)\:,
\end{align}
with solution
\begin{align}
\Phi(x) = -i \sum_\alpha Q_\alpha G_S(x-x_\alpha)\:,
\end{align}
where $G_S$ is the Mink$_3$ scalar $\Phi$ propagator. Therefore, Eq.~\eqref{eq:a-aBar-Solution}, reads 
\begin{align}
\widetilde{\mathcal{A}}_{u}(u, \bar{u}, z \rightarrow 0, t=0) 
= -\frac{g^2}{4} \sum_\alpha Q_\alpha \: \int \frac{d^2 u'}{2\pi} \: \frac{\partial_0 G_S(u' - u_\alpha, \bar{u}' - \bar{u}_\alpha, -t_\alpha)}{u-u'}\: .
\label{eq:A-solution-inTermsOfPropagator}
\end{align}

Let us compare this result with the holographic soft limit for non-simultaneous ${\cal O}_{\alpha}$, as given by
\begin{align}
j(u,\bar{u})= \sum_\alpha Q_\alpha \int \frac{d^2 u'}{u-u'}
\int \frac{dq_u}{2\pi} \, \frac{dq_{\bar{u}}}{2\pi} \: e^{i q_u (u_\alpha - u')}
\, e^{iq_{\bar{u}}(\bar{u}_\alpha - \bar{u}')}
\, e^{-2i\sqrt{q_u q_{\bar{u}}}t_\alpha}
\label{eq:j-in-terms-of-scalar-propagator}
\end{align}
following from Eqs.~\eqref{eq:j-non-simultaneous}, \eqref{eq:lambda-nonZeroTime}, \eqref{eq:lambda-zeroTime}. This precisely matches the form of memory, Eq.~\eqref{eq:A-solution-inTermsOfPropagator}, since the time-ordering in $G_S$ is fixed 
because all $t_{\alpha} < 0$.

The special case of $t_{\alpha} \rightarrow 0$ in AdS$_4$ is similar to the case of {\it massless} charges in Mink$_4$ reaching lightlike infinity, in each case leading to holomorphic $j$ with simple poles. We see this explicitly at $t_{\alpha} =0$ in Eq.~\eqref{eq:j-in-terms-of-scalar-propagator}, where the Fourier transforms give $\delta^2(u' - u_{\alpha})$. General $t_{\alpha} \neq 0$ in AdS$_4$ is similar to the case of {\it massive} charges in Mink$_4$ which approach timelike infinity,
in which case $j$ is not holomorphic. See Ref.~\cite{Strominger:2017zoo, Kapec:2015ena, Campiglia:2015qka} for the same smeared structure of poles in $\Mink_4$ memory for massive charges as our Eq.~\eqref{eq:a-aBar-Solution}. 
However, we see that in AdS$_4$  we have a clear holographic interpretation for this smearing in terms of the spreading of holographic charge density over time starting from $\delta$-function localization, $J_0 \propto \partial_0 G_S$, 
because the 3D charges are ``blobs" of massless CFT constituents. This is in contrast to 
 a 3D theory with only 3D-massive point-particle charges (without 4D dual), where $J_0$ would retain the form of $\delta$-functions  at particle locations over time, and the analogous construction of $j$ would have simple poles in $u$ without smearing over time. See the discussion in subsection~\ref{subsec:generalCFTstates}.

\section{Discussion}
\label{sec:discussion}
In this paper we have studied infinite-dimensional Kac-Moody (KM) asymptotic symmetries arising in $\AdS_4^{\Poincare}$ gauge theories. The standard asymptotic analysis, famously admitting only the finite-dimensional global symmetries of a holographically dual CFT$_3$,
was evaded in two steps, identified in Ref.~\cite{Mishra:2017zan} but taking their simplest form here.
In the present context, the major step was to consider alternate AdS boundary conditions peculiar to four dimensions, 
holographically dual to a modified $\widetilde{\CFT}_3$ obtained by an external Chern-Simons ($\CS$) gauging of the original $\CFT_3$. The second step
was to restrict attention to boundary/CFT correlators (or wavefunctional) at a fixed time, say $t=0$, where the  canonical CS structure yields holomorphic currents, whose Laurent expansion coefficients are KM charges. For more general correlators the physical essence of the KM symmetries is retained and generalized by the CS structure, but with a smearing out of the simple pole structure of KM holomorphic currents.
We showed how all this connects to ``holographic soft limits'' in AdS$_4^{\Poincare}$ which underlie  its KM asymptotic symmetries, for both abelian and non-abelian gauge fields. The 4D fields in this ``soft limit'' take the form of 3D CS fields (implying alternate boundary conditions for the AdS dual) which then lead to KM symmetries on an effectively 2D boundary of the CS spacetime, via the CS/WZW correspondence.

While soft limits yield the alternate/$\widetilde{\CFT}_3$ theory to leading order in the associated CS level, in the sense of Ref.~\cite{Mishra:2017zan}, it is interesting to see if the all-orders theory (finite CS level) can naturally emerge from the standard/$\CFT_3$ construction. We showed this for the case of {\it abelian} symmetry, where the standard construction imposed on electric-magnetic (mirror) dual variables assumes the alternate ($\widetilde{\CFT}_3$) form in the original variables, with finite CS level in the holographic description! The KM symmetries were thereby seen to be generalizations of dyonic charge conservation rather than simple electric charge conservation. It is less clear whether there is a non-abelian generalization, given the key role played by the S-duality transformation exchanging electric and magnetic charges. Perhaps a good theoretical laboratory is provided by those special supersymmetric non-abelian theories in which S-duality persists~\cite{Gaiotto:2008ak}.

There are several ways in which the KM structure derived in this work bears a resemblance to that of gauge theories in 4D Minkowski spacetime. It is useful to explore the similarities and differences in AS analyses of $\Mink_4$ and $\AdS_4$. In $\AdS_4$,  we have seen here and in Ref.~\cite{Mishra:2017zan} that holography allows us to straightforwardly and insightfully arrive at an AS structure previously unnoticed, whereas in $\Mink_4$ there is a  more familiar AS structure which may well point to some version of Minkowski holography, as yet unknown. In what follows, we comment on the similarities and differences, summarized briefly in Table~\ref{tab:mink-vs-ads}. The first hint that the AS structures in these two spacetimes may have some commonalities comes from the observation that the underlying CS gauge structure responsible for KM asymptotic symmetries in $\AdS_4$, was also seen in the Minkowski analysis of Ref.~\cite{Cheung:2016iub}. Yet, naively, a close resemblance would have seemed unlikely -- $\AdS_4$ and $\Mink_4$ are different spacetimes, with very different boundary structures. Further, $\Mink_4$ KM asymptotic symmetries reflect gauge-boson soft limits, whereas standard AdS$^{\rm global}$ lacks such soft limits. Nevertheless, we showed here that there is a simple generalization to ``holographic soft limits" in AdS$^{\Poincare}$ which underlies its KM asymptotic symmetries.

\renewcommand{\arraystretch}{1.5}
\begin{table}
\begin{tabular}{|p{7cm}|p{7cm}|}
\hline  
~~~~~~~~~~~~~~~~~~~~~~\textbf{Mink}$_4$ & ~~~~~~~~~~~~~~~~~~~~~~\textbf{AdS}$_4$ \\
\hline \hline
S-matrix & $\partial\AdS_4/\CFT_3$ local correlators \\
\hline
Timelike infinity $\equiv$ Euclidean $\AdS_3$  & $\partial\AdS_4 \equiv \Mink_3$ \\
\hline
Null infinity ($\mathcal{I}$), 2D geometry  & Fixed time $t=0$ on $\partial\AdS_4$, 2D geometry \\ 
\hline
Soft limit, $m_3\rightarrow 0$, where $m_3$ is the Casimir invariant of Euclidean $\AdS_3$~\cite{Cheung:2016iub} & Holographic soft limit, $m_3\rightarrow 0$, where $m_3$ is the Casimir invariant of $\Mink_3$   \\
\hline
$\CS$ structure of soft fields & $\CS$ structure of soft fields \\
\hline
2D holomorphic-WZW currents $j^a$ for (massless) charges hitting $\mathcal{I}$ & 2D holomorphic-WZW currents $j^a$ for charges hitting $t=0$ on $\partial\AdS_4$ \\ 
\hline
(Non-)abelian Kac-Moody AS & (Non-)abelian Kac-Moody AS \\
\hline
Electric/Magnetic Memories & Electric/Magnetic Memories \\
\hline
Electric flux Memory Kernel & Electric flux/Holographic charge density \\
\hline
~~~~~~~~~~~~~~~~~{\bf ?~?} & Holographic Duality \\
\hline
~~~~~~~~~~~~~~~~~{\bf ?~?} & $\widetilde{\CFT}_3$ with fully dynamical $\CS$ (finite level) 
\\
\hline
\end{tabular}
\caption{\small{The parallel developments between $\Mink_4$ and $\AdS_4$ gauge dynamics, their soft limits and associated 
infinite-dimensional KM asymptotic symmetries. AdS/CFT holography provides more of an explanatory structure in the case of AdS$_4$.}}
\label{tab:mink-vs-ads}
\end{table}

In Mink$_4$ gauge theory, massive charges emerging from a scattering event asymptotically approach future timelike infinity. This is a space parametrized by particle boosts, geometrically  3D hyperbolic space or, more suggestively, Euclidean $\AdS_3$~\cite{deBoer:2003vf}. Its boundary is future null infinity, ${\cal I}^+$, the destination for massless particles, which, while 3-dimensional, has 2D geometry due to the one null direction. In $\AdS_4$, there are also asymptotic 3D and 2D geometries. The asymptotic infinity of $\AdS_4^{\Poincare}$ is of course the boundary $\equiv \Mink_3$, the entire spacetime from the holographic perspective. The analogous ``2D boundary" for $\AdS_4$ is provided by a constant time slice on $\partial \AdS_4$, a boundary if one considers a wavefunctional on this time slice as determined by a path integral over just the earlier  spacetime region.
 Canonically in CS,  AS structure is associated to the wavefunctional, at say $t=0$, with its spacelike 2D geometry. In this work, we considered scattering in the bulk of $\AdS_4$, with some outgoing particles headed to the boundary and absorbed by local ($\CFT_3$) operators there. Charged particles arriving at $\partial \AdS_4$ at $t=0$ then play a somewhat  analogous role to massless charged particles arriving at ${\cal I}^+$ in Mink$_4$. This is seen more sharply by the soft CS/WZW structure that arises. In both cases, we get 2D holomorphic currents, with poles at the locations of the charges, and with Laurent expansions in terms of KM charges. Charged particles arriving at $\partial \AdS_4$ at more general $t \neq 0$ are the analogs of massive charges arriving at timelike infinity in $\Mink_4$ -- the 2D currents exist but are no longer holomorphic, the above-mentioned poles effectively being ``smeared"~\cite{Strominger:2017zoo, Kapec:2015ena, Campiglia:2015qka}. This smearing effect in the context of $\AdS_4$ finds a natural holographic explanation in the  tendency of 3D charge density to spread in a CFT$_3$ even if initially created in point-like form by a local operator. As discussed in subsection~\ref{subsec:generalCFTstates}, this smearing for general AdS$_4$ states does not compromise the KM structure, and furthermore in the CFT$_3$ dual description the smeared pole structure again resolves into discrete simple poles at the level of the 3D ``quarks" of the CFT.

The analogy between $\AdS_4$ and $\Mink_4$ is imperfect in one significant regard: while massless 4D charges robustly arrive at null infinity, ${\cal I}^+$, in $\Mink_4$, and massive charges do not, in $\AdS_4$ there is no such robust determinant of whether 4D charges will arrive at $\partial \AdS_4$ at $t=0$ or not. Instead, from the  3D Chern-Simons perspective the determining factor of whether  KM currents have robust simple pole structure or not is whether 3D charges are massive or massless, respectively. Of course, for the $\CFT_3$  dual to $\AdS_4$ the fundamental charges are massless.

Like in $\Mink_4$, $\AdS_4$ also has a close connection between KM symmetries and the memory effect, given by a large asymptotic spacetime Wilson loop. In $\AdS_4$, the analogous Wilson loop at the AdS boundary must vanish by standard boundary conditions. Nevertheless, we demonstrated that non-trivial ``magnetic" memory effects exist even with standard boundary conditions in $\AdS_4$, associated with non-vanishing 't Hooft loops on the boundary, and that these are closely related to holographic soft limits and KM structure.

It is an exciting open question as to how the rich structure of asymptotic symmetries and memories imply a new form of ``hair" for complex 4D states such as black holes, and can algebraically encode information that might seem lost according to standard 4D effective field theory analysis. We hope that the simple form and derivation of asymptotic symmetries and memories presented here  for $\AdS_4^{\rm Poincare}$, and the deep connection to holography, will help to answer this question in the future.

\acknowledgments
The research of RS was supported in part by the NSF under grant number PHY-1620074 and by the Maryland Center for Fundamental Physics (MCFP). The research of AM was supported in part by the NSF under grant number PHY-1407744. 

\bibliographystyle{JHEP}
\bibliography{References}

\end{document}